# Nanomagnetism in Otherwise Nonmagnetic Materials

Tatiana Makarova
Umeå University, Umeå, Sweden,
and Ioffe Physico-Technical Institute, St. Petersburg, Russia.
tatiana.makarova@physics.umu.se

> *For every complex problem there is an answer that
> is clear, simple, and wrong.*
> *H.L. Mencken*

**3. Ferromagnetism in hexaborides: discovery – disproof – rebuttal**

Young (Young 1999) reported ferromagnetism in La-doped calcium hexaboride ($CaB_6$) where a few of the calcium atoms a replaced with lanthium atoms. This discovery was taken as a mark of a long-sought mechanism for ferromagnetism in metals, where the 'electron gas' is susceptible to magnetic ordering at low density (Ceperley 1999); in other words, a phenomenon of high-temperature weak ferromagnetism at low-carrier concentration (HTFLCC) with no atomic localized moments. Later unusual ferromagnetism in hexaborides was reported for undoped $MB_6$ (M = Ca, Sr, Ba) (Ott 2000; Vonlanthen 2000). Sharp decrease in saturation with the increase of doping level ruled out the effects of an accident contamination but required consideration from the viewpoint of the electronic band structure. One school attributed this phenomenon to the polarization of low density electronic gas, another school to the hole doped excitonic insulator, whereas several authors insisted that the mechanism for magnetism in this compound is strongly connected to defects.

At the moment of the discovery of weak ferromagnetism alkaline-earth hexaborides were believed to be either semiconductors or semimetals (Massida 1997). Low electron doping with $La^{3+}$ (order of 0.1%) was shown to result in an itinerant type of ferromagnetism stable up to 600 K (Young 1999) and almost 1000 K. (Ott 2000). Magnetic order was ascribed to the itinerant charge carriers: a ferromagnetic phase of a dilute three-dimensional electron gas (Young 1999; Ceperley 1999). A different approach is based on the formation of excitons between electrons and holes in the overlap region around the point (Zhitomirskyi 1999, Murakami 2002)The idea of ferromagnetic instability in the excitonic metal was further developed by adding the effect of imperfect nesting on the excitonic state (Veillette 2002).

ESR experiments give some evidence that ferromagnetism of hexaborides is not a bulk phenomenon, but spins exist only within the surface layer of approximately 1.5 m thick (Kunii 2000). Band calculations (Jarlborg 2000) demonstrated a possibility for ferromagnetism below the Stoner limit. Observations of anomalous NMR spin-lattice relaxation for hexaborides showed the presence of a band with a coexistence of weakly interacting localized and extended electronic states (Gavilano 2001 ). Measurements of thermoelectric power and the thermal conductivity in the range of 5 – 300 K showed that it is described by scattering of electrons on acoustic phonons and ionized impurities in the conduction band, which is well separated from the valence band (Gianno 2002). The explanation of the origin of weak but stable ferromagnetism with the model of a band overlap has become a problem.

Parameter-free calculations of the single-particle excitation spectrum based on the so-called GW approximation predict a rather large band gap. $CaB_6$ should not be considered a semimetal but as a semiconductor with a band gap of 0.8 eV (Tromp 2001). Angle-resolved photoemission showed a gap between the valence and conduction bands in the X point, which exceeds 1 eV (Denlinger 2002). Assuming that $CaB_6$ is a semiconductor, magnetism is considered to be due to a La-induced impurity band, arising on the metallic side of the Mott transition for the impurity band (Tromp 2001).

Of alll intrinsic point defects, the B vacancy bears a magnetic moment of 0.04 $\mu_B$. The ordering of the moments can be understood in the assumption that in the presence of compensating cation vacancies a $B_6$ vacancy cannot be neutral (Monnier 2001). A support for the impurity-band mechanism has been found from the electrical and magnetic measurements on several La-doped samples (Terashima 2000).The models for a doped excitonic insulator included spatial inhomogeneity (Balents 2000) and phase separation with appearance of a superstructure (Barzykin 2000).

Shortly after the striking observation of high-temperature weak ferromagnetism, a discussion was opened concerning the possibility of a parasitic origin of the hexaboride ferromagnetism (Matsubayashi 2002). Replying to the claims, the authors of the pioneering work (Young 2002) noted that these new findings did not contradict their picture (Fisk 2002) where the magnetism is due to strongly interacting magnetically active defects in off-stoichiometric $CaB_6$ crystals. Iron with a concentration of about 0.1 atomic per cent is indeed involved in the weak high-temperature ferromagnetism of $CaB_6$ although the exact mechanism is still unclear and probably highly nontrivial (Young 2002). Sources of iron were boride commercial powders, boron powder, and aluminum metal, used in significant quantity as a flux for crystal growth (Otani 2002; Otani 2003). Iron is concentrated in the surface region (Meegoda 2003). Fe and Ni are found at the edges of facets and growth steps (Bennett 2004

The understanding that iron is somehow involved in the ferromagnetism of $CaB_6$ has not stopped claims the magnetism is of a nontrivial origin. The presence of transition metals could not account for the dependence of magnetism on La or Ca concentration (Young 2002, Cho 2004), enhanced magnetization and increased ordering temperature during the sample storage in air (Lofland 2003). No dependence of the saturation magnetization on Fe concentration was found (Young 2002). Having studied $CaB_6$ crystals of different purity, the experimenters analyzed the formation of the mid-gap states and suggested that the exotic ferromagnetism in $CaB_6$, *in part,* cannot be ascribed to a magnetic impurity (Cho 2004).The experiments on thin films (Dorneles 2004) have shown that Fe/Ca atomic ratio exceeding 100% would be needed to explain the sample magnetization. Huge magnetic moments were detected at the surface layers of thin films of disordered $CaB_6$ and $SrB_6$ (Dorneles 2004). Lattice defects are suggested as the origin of the high-temperature magnetism in hexaborides.

*Ab initio* calculations (Maiti 2008) show that a vacancy exhibits a magnetic moment. A photoemission study (Rhyee 2004) reveals the presence of weakly localized states in the vicinity of the Fermi level in the ferromagnetic $CaB_6$, whereas it is absent in the paramagnetic $LaB_6$ (Medicherla 2007). All these results suggest that the nature of ferromagnetism is intrinsic. Its nature is different from that of diluted magnetic semiconductors because it is *related to the presence of localized states at the Fermi level* originating from the boron vacancies. The model (Edwards and Katsnelson 2006) has been able to explain not only ferromagnetism of undoped $CaB_6$, but also sheds light on the non-trivial role of iron in iron-contaminated samples.

## 4. Magnetic semiconducting oxides

Dilute magnetic oxides (DMO) is another example of systems where magnetism appears *eh nihilo*. As predicted by Dietl et al.,(Dietl 2000) ZnO and GaN can be the host candidate for the room-temperature ferromagnetic DMS. There is growing evidence that the mechanism responsible for the emergent behaviour is different from DMS and stems from the interface effects (Brinkman 2006, Hernando 2006), although the carrier mediation mechanism is still under discussion (Calderon 2007).

**4.1 Zink oxide** First observations of magnetism in DMO were made on the oxides $TiO_2$ (Matsumoto 2001) and ZnO (Ueda 2001) doped with a transition metal cobalt. Question of intrinsic nature arose (Norton 2003), since some studies reported the phase separation and the formation of ferromagnetic clusters (see Janish 2005 for a review).

The most recent results have demonstrated that oxide thin films or nanostructures do not need magnetic cations to become magnetic Cobalt *suppresses* ferromagnetism in ZnO (Ghoshai 2008). RTFM was found in non-transition-metal-doped ZnO (Coey 2005c), in ZnO single crystals implanted by Ar ions (Borges 2007), doped with carbon (Pan 2007). Enhancement of RTFM upon thermal annealing was found in pure ZnO (Banerjee 2007) and explained by the formation of the anionic vacancy clusters.

RTFM was attributed to Zinc vacancies (Xu 2008), (Wang 2008). A robust ferromagnetic state is predicted at the O (0001) surface (Sanches 2008). ZnO nanowires show RTFM (Yi 2008) owing to Zn clusters. ZnO films covered with Zn, Al, Pt do show RTFM while (Ag, Au)/ZnO do not (Ma 2008). ZnO becomes ferromagnetic once oxygen defects are introduced in it (Sanyal 2008). The FM in this type of compound stem from defects on Zn sites (Hong 2007b). Size and shape of nanocrystals are important. (Yan 2008). Ferromagnetic order can be induced in ZnO by 2p light element (N) doping (Shen 2008) or by means of Fe ion implantation or just by vacuum annealing at mild temperatures without any transition metal doping (Zhou 2008). The model of oxygen vacancies (Chakraborti 2007) contradicts to the DFT calculations (Ye 2006) which show that vacancies tend to destroy ferromagnetism. The experiments on ZnO:Cu nanowires (Shuai 2008) speak against the oxygen vacancies. FM may occur due to the hybridization between partly occupied Cu $3d$ bands and O $2p$ bands: these energy levels are closely situated, and the delocalized holes induced by O $2p$ and Cu $3d$ hybridization can efficiently mediate the ferromagnetic exchange interaction (Huang 2006). XMCD studies on Co and Li doped ZnO haven't revealed *any element specific signature of ferromagnetism.* . This result suggest that RTFM in doped ZnO has an intrinsic origin and is caused by the oxygen *vacancies* (Tietze 2008). Ferromagnetism in Ga-doped ZnO was attributed to vacancies (Bhosle 2008), the Zn vacancy (Wang 2008). (Potzder 2008) used hammer for inducing RTFM.

RTFM in Cu-doped ZnO was theoretically predicted (Ye 2006, Park 2003, Wu 2006) and observed experimentally (Ando 2001, Cho 2004, Chakraborti 2007, Hou 2007, Xing 2008) . However, a magnetic circular dichroism study in Cu-doped ZnO thin films did not detect any significant spin polarization on the Cu $3d$ and $O_2 p$ states, although the samples showed RTFM and were free of contamination (Keavney 2007). Several Cu-doped oxides have been studied (Dutta 2008), but only ZnO oxide demonstrates ferromagnetic behaviour, thus the CuO phase is suggested to be a paramagnet. A comparative study on Cu-doped ZnO nanowires prepared by two distinct methods demonstrated an enhancement of RTFM by structural inhomogeneity (Xing 2008). The results suggest that RTFM is not a homogeneous bulk property, but a *surface effect by nature.*

**4.2. Titanium oxide** A semiconducting material, $TiO_{2-\delta}$ is ferromagnetic up to 880 K, without the introduction of magnetic ions (Yoon 2006). The subscript (2-δ) implies oxygen deficiency in the samples, or the presence of oxygen vacancies. RTFM has been observed in $TiO_{2-\delta}$ nanoparticles synthesized by the sol-gel method and annealed under different reducing atmosphere (Zhao 2008). (Zhao 2008) explain their results by the aggregation of the oxygen vacancies. Pure $TiO_2$ thin films demonstrated RTFM when annealed in vacuum (Sudakar 2008), while the air-annealed samples showed negligible magnetic moments.

Spin coated pristine $TiO_2$ thin films show magnetic behavior similar to that of pulsed laser ablated TiO2 thin films (Hassini 2008). RTFM is induced by oxygen ion irradiation (Zhou2009). The origin of the collective behavior is ascribed to magnetic moments is the holes introduced by the <u>titanium vacancies but not oxygen vacancies</u> (Peng2009) The set of experimental data on transition metal-free oxides raised the question: Does Mn doping play any key role in tailoring the ferromagnetic ordering of $TiO_2$ thin films? (Hong 2006), When the Mn concentration is small and does not distort the $TiO_2$ structure, it enhances the ferromagnetic component into the magnetic moment of the already ferromagnetic $TiO_2$ film base, which is already ferromagnetic, then enhances it. But at higher concentrations Mn drastically degrades and destroys the ferromagnetic ordering. These experiments show that Mn doping indeed does not play any key role in introducing FM in $TiO_2$ thin films. In other words, magnetic semiconducting oxides cannot be considered as a variety of dilute magnetic semiconductors DMS, *DMO are not DMS.*

**4.3. Hafnium oxide.** Unexpected magnetism was observed in undoped $HfO_2$ thin films on sapphire or silicon substrates (Venkatesan 2004). While for the films prepared by pulsed laser deposition the results were confirmed (Hong 2006), other groups did not observe the effect on the films grown by metallorganic chemical vapor deposition (Abraham 2005) and by pulsed-laser deposition (Rao 2006). Weak signal in lightly Co-doped $HfO_2$ films it came presumably from a Co rich surface layer (Rao 2006). The ferromagnetic signal was ascribed to possible tweezer contamination (Hadacek 2007). An intrinsic magnetism was not identified either on films, or on $HfO_2$ powders annealed in pure hydrogen flow (Wang 2006). On the theoretical side, the RTFM in the $HfO_2$ system is not forbidden. Isolated cation vacancies in $HfO_2$ may form high-spin defect states resulting in a ferromagnetic ground state (Das Pemmaraju 2005). A model for vacancy-induced ferromagnetism is based on a correlated model for oxygen orbitals with random potentials representing cation vacancies (Bouzerar 2006). Colloidal $HfO_2$ nanorods with controllable defects have been synthesized (Tirosh 2007). Defects have been studied by high-resolution electron microscopy and by optical absorption spectroscopy, and it was shown that nanocrystals with a high defect concentration exhibit ferromagnetism and superparamagnetic-like behavior. The RTFM of amorphous $HfAlO_x$ thin films has been demonstrated (Qiu 2006) and interfacial defects suggested as are one of the possible sources .

**4.4. Other nonmagnetic oxides.** Ferromagnetism in oxides can be induced by the elements of Periodic Table which have nothing to do with magnetism in the bulk state. *Ab initio* study (Maca 2008) of the induced magnetism in $ZrO_2$ shows that the substitution of the cation by an impurity from the groups IA or IIA of the Periodic Table (K and Ca) leads to opposite results: K impurity induces magnetic moment on the surrounding O atoms in the cubic $ZrO_2$ host whilst Ca impurity leads to a nonmagnetic ground state. Similarly, experiments on Mn-doped $SnO_2$ films show that a transition-metal doping does not play any key role in introducing FM in the system (Hong 2008). Observation of room temperature ferromagnetism in nanoparticles of nonmagnetic oxides such as as $CeO_2$, $Al_2O_3$, ZnO, $In_2O_3$ and $SnO_2$ (Sundaresan 2006) lead to a conclusion which is probably adventurous: ferromagnetism is a universal feature of nanoparticles of otherwise non-magnetic oxides. $CeO_2$ nanoparticles and nanocubes have been investigated both experimentally and theoretically (Ge 2008), and it is found that monodisperse $CeO_2$ nanocubes with an average size of 5.3 nm do show ferromagnetic behavior at ambient temperature. Size dependent ferromagnetism in cerium oxide nanostructures was found to be independent of oxygen vacancies (Liu 2008). Ferromagnetism in nanosized $CeO_2$ powders was studied in nanoparticles of different sizes but found only in sub-20 nm powders. Remarkable room-temperature ferromagnetism was observed in undoped $TiO_2$, $HfO_2$, and $In_2O_3$ thin films (Hong 2006). In another study, no trace of ferromagnetism has been detected in $In_2O_3$ (Berardan 2008). Room temperature size dependent ferromagnetism was observed in sub-20 nm sized $CeO_2$ nanopowders; FM <u>is not linked to oxygen</u> vacancies, but possibly to the changes of a cation surface defect state. The occurrence of spin polarization at $ZrO_2$, $Al_2O_3$ and MgO surfaces is proved by means of ab initio calculations within the density functional theory (Gallego 2005).

Creation of collective ferromagnetism in nonmagnetic oxides by intrinsic point defects such as vacancies has been discussed (Osorio-Gullien 2007).
.

**5. Magnetism in metal nanoparticles**
Magnetism in ~3 nm gold nanoparticles (Au NPs) with an unexpected large magnetic moment of about 20 spins per particle (Hori 1999) was followed by several papers with the emphasis on the stabilization of the particles by various polymers, diameter dependence of the ferromagnetic spin moment, experimentally (Hori 2004) and theoretically (Michael 2007) and observation of spin polarization of gold by x-ray magnetic circular dichroism.

Self-assembled alkanethiol monolayers on gold surfaces have been reported to show permanent magnetism (Crespo 2004). Au NPs capped with dodecanethiol showed superparamagnetic or diamagnetic behaviour depending on ots size. (Dutta 2007). It has been suggested that ferromagnetism is associated with 5d localized holes generated through Au-S bonds (Crespo 2004). According to electron circular dichroism measurements carried out on thiolated organic monolayers on gold (Vager 2004), the magnetic moment originates from the orbital momentum. Highly anisotropic giant moments were also observed for self-organized organic molecules linked by thiols bonds to gold films (Carmelli 2003). The magnetic moment reaches 10 or even 100 $\mu_B$ per atom) for films, but it is extremely low (0:01 $_B$ per atom) for nanoparticles. Probably, this phenomenon is due to the directional nature of the assembled organic layers. Self-assembled monolayers on gold of double-stranded DNA oligomers create a strong and oriented magnetic field (Ray 2006). An explanation invoking strong spin-orbit interaction is given(De La Venta 2007).

X-Ray magnetic circular dichroism experiments (Yamamoto 2004) provided direct evidence for ferromagnetic spin polarization of Au nanoparticles. Further XMCD studies (Negishi 2006) confirmed that the presence of localized holes created by Au-S bonding at the interface. The experimens with different capping agents clarified the role of adsorbing molecules (Guerrero 2008). Reversible phototuning of ferromagnetism was observed in gold nanoparticles (Suda 2008). A direct observation of the intrinsic magnetism of Au-atoms in thiol-capped gold nanoparticles, which possess a permanent magnetisation at room temperature, (Garitaonandia 2008) was obtained by using two element specific techniques: X-ray magnetic circular dichroism on the L edges of the Au and Au-197 Moessbauer spectroscopy.

Some 4d elements, nominally nonmagnetic, e.g., Ru, Rh, Pd, (Sampedro 2003), (Ito 2008). may also exhibit nanoscale magnetism. Platinum atoms, not magnetic in the bulk, become magnetic when grouped together in small clusters (Liu 2006). Platinum monatomic nanowires were predicted to spontaneously develop magnetism, (Smogunov 2008b), and it was shown that Pd and Pt nanowires are ferromagnetic at room temperature, in contrast to their bulk form (Teng 2008). Magnetic properties of Ru/Ta mixture drastically changes by Xe atom irradiation, and the reason is the formation of small clusters in the Ru/Ta matrix under irradiation (Wang 2006). Potassium clusters display a non-trivial magnetic behaviour on the nanoscale: low T ferromagnetism when the clusters are incorporated into zeolite lattice (Nozue 1992). Two models have been invoked to explain this beghaviour: spin-canting mechanism of antiferromagnet (Nakano 2000 and N-type ferrimagnetism (Nakano 2006) which is constructed of non-equivalent magnetic-sublattices of K clusters the matrix. Accommodated in the magnetic nanographene-based porous network potassium clusters, which are 60 atoms on average, become antiferromagnetic (Takai 2008) due to the charge transfer with the host nanographene.

The results on metal films and nanoparticles point out the possibility to observe magnetism at nanoscale in materials without transition metals and rare earths atoms, and are of fundamental value to understand the magnetic properties of surfaces. The available explanation of orbital ferromagnetism and giant magnetic anisotropy at the nanoscale (Hernando 2006) assumes the induction of orbital motion of surface electrons around ordered arrays of Au–S bonds

## 6. Magnetism in semiconductor nanostructures

Semiconductor nanoparticles show increasing magnetization for decreasing diameter (Neeleshwar 2005). RTFM in undoped GaN and CdS semiconductor nanoparticles of different sizes was observed for the particles with the average diameter in the range 10-25 nm. RT saturation magnetization is of the order of $10^{-3}$ emu/g, which is comparable to that observed in nanoparticles of nonmagnetic oxides. . Agglomerated particles of GaN and CdS loose the FM properties: the saturation magnetic moment decreases with the increase in particles size, suggesting that ferromagnetism is due to the defects confined to the surface of the nanoparticles (Madhu 2008). Ferromagnetism has been also measured in PbS attached to the GaAs substrate (Zakrassov 2008). RTFM in CdSe quantum dots (QD) capped with TOPO (tri-n-octylphosphinehas) been observed (Seehra 2008). The strength of magnetism weakens with increase in size of the QDs. This phenomenon is classified as *ex-nihilo* magnetism since the effect stems from the contact of two diamagnetic materials, namely CdSe and TOPO. The magnetism here is possibly due to the charge transfer from Cd d-band to the oxygen atoms of TOPO.

Adsorption of monolayers of organic molecules onto the surface of ferromagnetic semiconductor heterostructures produces large, robust changes in their magnetic properties (Kreutz 2003). Effects of chemisorption of polar organic molecules onto ferromagnetic GaAs/GaMnAs heterostructures has been investigated (Carmeli 2006). The new electronic and magnetic properties emerge from the charge transfer from the organized organic layer substrate and, in particular, from the alignment of the spin of the transferred electrons/holes (Naaman 2006). Defect-induced ma GaN is believed to be due to a Ga vacancy defect which can show induced local magnetic moment in N atoms (Hong J 2008). Cation-vacancy induced intrinsic magnetism in GaN and BN is investigated, and a dual role of defects is shown: First, the defects create a net magnetic moment, and second, the extended tails of defect wave functions mediate surprisingly long-range magnetic interactions between the defect-induced moments (Dev 2008).

### 7. Ferromagnetism in carbon nanostructures

**7. 1. Pyrolitic carbonaceous materials.** Magnetic ordering at high temperatures in carbon-based compounds has been persistently reported since 1986 (reviewed in Makarova 2004). The earlier results on magnetic carbon compounds obtained by pyrolysis of organic compounds at relatively low temperatures were poorly reproducible.

**7.2. Graphite.** Ferromagnetic and superconducting-like magnetization hysteresis loops in highly oriented pyrolitic graphite (HOPG) samples above room temperature were reported (Kopelevich 2000). Later the authors retracted from superconducting-like hysteresis loops as they had been partially influenced by an artefact produced by the SQUID current supply, but ferromagnetic behaviour remained beyond doubts (Kopelevich and Esquinazi 2007). Absence of correlation between magnetic properties and impurity content was found in highly oriented pyrolytic graphite (Esquinazi 2002), suggesting intrinsic ferromagnetic signal.

**7.3 Porous graphite.** Various independent groups have reported ferromagnetism and anomalous magnetic behaviours in porous graphitic based materials. Ferromagnetic correlations have been observed in activated mesocarbon microbeads mainly composed of graphitic microcrystallites (Ishii 1995). RTFM has been found in microporous carbon with a zeolite structure (Kopelevich 2003). Similar behaviour was described for glassy carbon (Wang 2002), The carbon nanofoam (Rode 2004) exhibits ferromagnetic-like behaviour with a narrow hysteresis curve and a high saturation magnetization. Strong magnetic properties fade within hours at room temperature; however, at 90 K the foam's magnetism persists for up to 12 months. Magnetic behaviour is close to the spin glass picture with unusually high freezing temperature (Blinc 2006). The higher *g* factors are typical of amorphous carbon systems with strongly nonplanar parts of a carbon sheet (Arcon 2006). Oxygen eroded graphite (Mombru 2005, Pardo 2006) shows multilevel ferromagnetic behaviour with the Curie temperature at about 350 K.

**7.4. Carbon nanoparticles.** For carbon nanoparticles prepared in helium plasma (Akutsu 1999), the saturation magnetization increases with decreasing grain size, and the grain size of the carbon fine particles having the highest magnetization is 19 nm. More recently, ferromagnetism was found in carbon nanospheres (Caudillo 2006), macrotubes (Li 2007) necklace-like chains and nanorods (Parkansky, 2008), highly oriented pyrolytic graphite nanospheres grown from Pb-C nanocomposites (Li 2008). Interestingly, magnetic carbon nanoparticles have mainly spherical shapes.

**7.5. Nanographite.** Various types of magnetic behaviour were discovered in nanocarbon derived from graphite (Andersson 1998). Strong antiferromagnetic coupling has been found between the spins localized on the surface of similar particles (Osipov 2006). Another example is activated carbon fibers (Shibayama 2000 indicating a presence of a quenched disordered magnetic structure like a spin glass state. Physisorptoiuon of water drastically changes magnetic properties, although water itself is nonmagnetic (Sato 2003). Water molecules compress the nanographite domains, reducing the interlayer distance in a stepwise manner. (Sato 2007). The physisorption of various guest materials can cause a reversible low-spin/high-spin magnetic switching phenomenon, while oxyhen molecules physisorption is responsible for the giant magnetoresistance of the nanographite network (Enoki 2008).

**7.6. Fullerenes.** Polymer – $C_{60}$ composite with room temperature ferromagnetism was first reported when $C_{60}$ was ultrasonically dispersed in a dimethylformamide solution of polyvinylidenefluoride (Ata 1994). Fullerene hydride $C_{60}H_{36}$ (Lobach 1998) and $C_{60}H_{24}$ (Antonov 2002) have been reported to be room-temperature ferromagnets. Room-temperature ferromagnetism of polymerized fullerenes was first reported when the samples were exposed to oxygen under the action of the strong visible light (Murakami 1996). In photopolymers saturation magnetization progressively increases with increasing exposure time (Makarova 2003). The existence of ferromagnetic phase in photolyzed $C_{60}$ was confirmed by the three methods. (1) SQUID; (2) Ferromagnetic resonance in the EPR; (3) Low-field nonresonance derivative signal (Owens 2004). The experiments were done in a chamber with flowing oxygen, which exclude any possibility of penetration of metallic particles during the experiment. The presence of a magnetically ordered phase was revealed in pressure-polymerized $C_{60}$ (Makarova 2001). Later several authors retracted from this paper on the grounds that the impurity content measured on the surface of the samples by Particle Induced X-Ray scattering (PIXE) was higher than that obtained by the bulk impurity analysis and that the Curie temperature was close to that of $Fe_3C$. Some authors did not agree with the retraction. One of the reasons of disagreement was that the concentration of impurities within the information depth of the PIXE method (36 μm) measured on the same samples was still 3 times less than necessary for the observed magnetic signal (Han 2003, Spemann 2003). The whole set of experiments was repeated with the same team of technologists and on the same equipment (Makarova and Zakharova 2008), and for several samples the magnetization values were higher than those expected from the metallic contamination. Having followed in situ the depolymerization process through the temperature dependence of the ESR signal (Zorko 2005), the authors conclude that the magnetic signal is directly connected with the polymerized fullerene phase and cannot be attributed to iron compounds. Systematic study of synthesis conditions for the production of the ferromagnetic fullerene phase was made by another team. Only samples prepared in a narrow temperature range show a ferromagnetic signal with a qualitatively similar magnetic behavior (Narozhnyi 2003). A different method was used for the preparation of the ferromagnetic polymers of $C_{60}$: multi-anvil octupole press (Wood 2002). Inelastic neutron scattering analysis of the ferromagnetic phase in the polymerized fullerene sample showed a sufficient presence of hydrogen (Chan 2004). $C_{60}$ adsorbed on silicon surface has an antiferromagnetic ground state (Lee 2009).

### 7.7. Irradiated carbon structures

Studies of irradiated carbon structures provided convincing proof for the intrinsic origin of the effect. This is the case of the proton-irradiated HOPG where the ultimate purity of the material is proved by simultaneous measurements of the magnetic impurities (Esquinazi 2003).Elementally sensitive experiments on proton bombarded graphite provided fast evidence for metal-free carbon magnetism (Ohldag 2007). The temperature behaviour suggests two-dimensional magnetic order (Barzola 2007). Ferromagnetism was found in irradiated fullerenes with 250 keV Ar and 92 MeV Si ions, (Kumar 2006), with 10 MeV oxygen ion beam (Kumar 2007), 2 MeV protons (Mathew 2007). Paradoxically, if one bombards graphite with iron and hydrogen, both produce similar paramagnetic contributions. However, only protons induce ferromagnetism (Hohne 2008, Barzola 2008).

Mechanism for ferromagnetism in $H^+$ irradiated graphite is largely unknown and may result from the appearance of bound states due to disorder and the enhancement of the density of states (Araújo 2006), can be induced by single carbon vacancies in a three-dimensional graphitic network (Faccio 2008); magnetism decreases for both diamond and graphite with increase in vacancy density (Zhang 2007). The role of hydrogen is not well understood as magnetism should survive only at low H concentrations (Boukhvalov 2008). The mechanism of ferromagnetism in disordered graphite samples is considered to arise from unpaired spins at defects, induced by a change in the coordination of the carbon atoms (Guinea 2006). Several works discuss theoretical models which address the effects of electron–electron interactions and disorder in graphene planes (González 2001, Stauber 2005).

### 7.8. Magnetic nature of intrinsic carbon defects

Defects in graphite always reduce the diamagnetic signal. In a simplified picture, vacancies, adatoms, pores and bond rotations enhance local paramagnetic ring currents and produce local magnetic moments (Lopez-Urias 2006). Theory allows also magnetism in diamond structures. (Cho 2008).

**Adatoms and vacancies.** Carbon adatoms possess a magnetic moment of about 0.5 μB whereas carbon vacancies in graphitic network generate a magnetic moment of about 1 μB (Ma 2004). Vacancies in graphite, both ordinary and hydrogenated create new states below the Fermi level. (Lehtinen 2003)

**First-raw elements** A specific case of defects is the presence of the first-raw elements, although they cannot be unambiguously classified as intrinsic carbon defects. The most important defect is hydrogen: Unsaturated valence bonds at the boundaries of graphene flakes are filled with stabilizing elements; among these stabilizers hydrogen atoms are the common ones. The entrapment of hydrogen by dangling bonds at the nanographite perimeter can induce a finite magnetization. A theoretical study of a graphene ribbon in which each carbon atom is bonded to two hydrogen atoms at one edge and to a single hydrogen atom at the other edge shows that the structure has a finite total magnetic moment (Kusakabe 2003). Combination of different edge structures (by means of hydrogenation, fluorination or oxidation) is proposed as a guiding principle to design magnetic nanographite (Maruyama 2004). Hydrogenation of carbon materials can induce magnetism through termination of nanographite ribbons, adsorption on the CNT external surface (Pei 2006), trapping at a carbon vacancy or pinning by a carbon adatom (Ma 2005).

Other elements which may strongly influence magnetic behaviour of carbon are boron and nitrogen. Border states in hexagonally bonded BNC heterosheets have been predicted to lead to a ferromagnetic ground state, a manifestation of flatband ferromagnetism (Okada and Oshiyama 2001). In heterostructured nanotubes, partly filled states at the interface of carbon and boron nitride segments may acquire a permanent magnetic moment. Depending on the atomic arrangement, heterostructured C/BN nanotubes may exhibit an itinerant ferromagnetic behavior owing to the presence of localized states at the zigzag boundary of carbon and boron nitride segments (Choi 2003)

**Curvature** Stone-Wales defects and related structures with negative Gaussian curvature. Gaussian negative curvature provides a mechanism for steric protection of the unpaired spin (Park 2003). A particular case of a graphene modification is the Stone-Wales defects, or topological defects, caused by the rotation of carbon atoms which leads to the formation of five- or sevenfold rings. A novel class of curved carbon structures, Schwarzites and Haeckelites, has been proposed theoretically (Mackay 1991). Schwarzite is a form of carbon containing graphite-like sheets with hyperbolic curvature, So far, periodic schwarzites have not been realized experimentally; however, there is experimental evidence that random Schwarzite structures are present in a cluster form in such carbon phases as spongy carbon (Barborini 2002) and carbon nanofoam (Rode 2004). In the systems with negative Gaussian curvature, unpaired spins can be introduced by sterically protected carbon radicals. Not only negative Gaussian curvature may lead to magnetism; the same is true for the positive: carbon compounds that display an odd number of pentagons and heptagons, present polarization in the ground state (Azevedo 2008).

**Zigzag edges** Nanographite is characterized by the dependence of electronic structure on edge termination: edge states are present on variously terminated zigzag edges but are absent at the armchair edges (Fujita 1996). These states produce large electronic density of states at the Fermi level and play an important role in the unconventional nano-magnetism (Enoki 2007). It is suggested that the basic magnetic mechanism is spin polarization in these highly degenerate orbitals or in a flat band (Kusakabe 2006). Similar predictions have been made for ZnO nanoribbons with zigzag-terminated edges (Botello-Mendez 2008). Zigzag carbon nanotubes may exhibit different types of ordering (Wu2009).

**Defects in fullerenes** Ferromagnetism in fullerenes is also thought to be of defect nature. It was shown both theoretically (Okada and Oshiyama 2003) and experimentally (Boukhvalov 2004) that the ideal polymerized fullerene matrix is not magnetic. Partial disruption of interfullerene bonds: linking of molecules through a single bond is preferred for multiplet states of system (Chan 2004). Cage distortion of $C_{60}$ in polymerized two-dimensional network leads to competition between diamagnetic and ferromagnetic states (Nakano 2004). Certain types of vacancies in coexistence with the 2 + 2 cycloaddition bonds represent a generalized McConnell's model for high-spin ground states in the systems with mixed donor-acceptor stacks (Andriotis 2005). Donor-acceptor mechanism was considered for the case of microscopic electric charge inhomogeneities introduced in a polymeric network. Two charged adjacent fullerenes interact ferromagnetically, and the ground state of a charged dimer is triplet (Kvyatkovskii 2004). The $C_{60}$ doublet radicals appear after the application of pressure, and this state has a long life state (Ribas-Arino 2004a). The evaluation of capability of the $C_{60}$ molecule to act as a magnetic coupling unit was made: $C_{60}$ diradical is an excellent magnetic coupler (Ribas-Arino 2004b). Some metastable isomer states with zigzag-type arrangement of the edge atoms of $C_{60}$ may form during the cage opening process (Kim 2003). Long-range spin coupling, which is an essential condition for the ferromagnetism, has been considered through the investigation of an infinite, periodic system of polymerized $C_{60}$ network. Chemically bonded hydrogen plays a vital role, providing a necessary pathway for the ferromagnetic coupling of the considered defect structure. $C_{60}$ molecules become magnetically active due to the spin (and charge) transfer from dopants. Magnetic transitions were reported for the TDAE-$C_{60}$ ferromagnet, the $(NH_3)K_3C_{60}$ antiferromagnet, $AC_{60}$ and $Na_2AC_{60}$ polymers (A = K, Rb, Cs). Refs. (Kvyatkovskii 2005, Kvyatkovskii 2006) consider the situation when $C_{60}$ molecule is doped through the presence of structural defects and impurities which create stable molecular ions $C_{60}^{\pm}$ and analyze the interaction of two adjacent molecules (i.e. dimer) embedded in a two-dimensional polymeric network. The main result is that the ferromagnetic interaction is possible only in the crystals where fullerenes have specific orientation. This type of orientation is provided by (2+2) cycloaddition reaction which forms a double bond (DB) between the buckyballs.

### 7.9. Magnetism of graphene

Experimentally, the observation of room-temperature graphene magnetism was claimed on the graphene material prepared from graphite oxide (Wang 2009).

It has been shown theoretically (Vozmediano 2006) that the interplay of disorder and interactions in a 2D graphene layer gives rise to a rich phase diagram where strong coupling phases can become stable. The theories predict itinerant magnetism in graphene due to the defect-induced extended states (Yazev and Helm 2007) while only *single-atom defects* can induce FM in graphene-based materials (Yazev 2008) or short-range magnetic order peculiar to the honeycomb lattice with the vacancies (Kumazaki 2007a) or with hydrogen termination or a chemisorption defect (Kumazaki 2007b). The graphene magnetic susceptibility is temperature dependent, unlike an ordinary metal (Kumazaki 2007c). Spin susceptibility which decreases with temperature without impurities, takes a finite value with impurities which may enhance the tendency to a ferromagnetic ordered state. (Peres 2006).

Finite graphene fragments of certain shapes, e, g, triangular or and hexagonal "nanoislands" terminated by zigzag edges (Fernandez-Rossier and Palacios 2007), or varable-shaped graphene nanoflakes (Wang 2008), as well as some "Star of David"-like fractal structures (Yazev 2008) possess a high-spin ground state and behave as artificial ferrimagnetic atoms. Ferrimagnetic order emerges in rhombohedral voids with imbalance charge in graphene ribbons, and the defective graphene ribbons behave as diluted magnetic semiconductors (Palacios and Fernández-Rossier 2008). A defective graphene phase is foreseen to behave as a room temperature ferromagnetic semiconductor (Pisani). Both magnetic and ferroelectric orders are predicted (Fernandez-Rossier 2008).

Edge state magnetism has been studied on realistic edges of graphene and is shown that only elimination of of zigzag parts with n > 3 will suppress local edge magnetism of graphene (Kumazaki 2008). The edge irregularities and defects of the bounding edges of graphene nanostructures do not destroy the edge state magnetism (Bhowmick 2008). However, such edge defects (vacancies) and impurities (substitutional dopants) suppress spin polarization on graphene nanoribbons which is caused by the reduction and removal of edge states at the Fermi energy (Huang 2008). Magnetic order in zigzag bilayers ribbons is also related to the properties of zigzag edges (Sahu 2008). Neutral graphene bilayers are proposed to be pseudospin magnets (Min 2008). In a biased bilayer graphite (Stauber 2008) a tendency towards a ferromagnetic ground state is investigated and shown that the phase transition between paramagnetic and ferromagnetic phases is of the first order. Spin is confined in the superlattices of graphene ribbons, and in specific geometries magnetic ground state changes from antiferromagnetic to ferrimagnetic (Topsakal 2008).

The superlattices of graphene nanoholes exhibit long-range magnetic
order and collective "bulk" magnetism (Yu2008).

An alternative approach is connected with pentagons, dislocations and other topological defects (Carpio 2008). Single pentagons and glide dislocations made of a pentagon-heptagon pair alter the magnetic behaviour, whereas the Stone-Wales defects are harmless in the flat lattice.

The combination of hydrogen-induced magnetism and changeable thermodynamics
upon variation of the graphene layer spacing makes graphene a reversible magnetic system
(Lei 2008). In graphene magnetism survives at low H concentrations (Boukhvalov 2008). A number of nanoscale spintronics devices utilizing the phenomenon of spin polarization localized at one-dimensional (1D) zigzag edges of graphene have been proposed (Yazyev and Katznelson 2008). Some of the theories created for graphene explain the experimental observations in observations proton bombarded graphite (Yazyev 2008b).

A DFT study of single and double vacancies (SV and DV) in a graphene shee showed that metals embedded in the vacancies exhibit interesting magnetic behavior. In particular, an Fe atom on a SV is not magnetic, while the Fe@DV complex has a high magnetic moment. Surprisingly, Au and Cu atoms at SV are magnetic (Krasheninnikov 2009). Magnetism at graphene edges is fragile with respect to oxidation, on the other hand, hydrogenation of the Stone-Wales defects may be a prospective way to create magnetic carbon (Boukhvalov 2008).

Carbon materials that exhibit ferromagnetic behaviour have been predicted theoretically and reported experimentally in recent years (Makarova and Palacio 2006). The initial surprising experiments were confirmed by the independent groups. The fact that carbon atoms can be magnetically ordered at room temperature was confirmed by the direct experiment: an element-sensitive method X-ray magnetic circular dichroism. Electron spins in diamond show the longest room-temperature spin dephasing times ever observed in solid-state systems which makes them very promicing for spintronics (Balasubramanian2009).

## 8. Possible traps in search of magnetic order

The source of RTFM is often controversial and contentious particularly because of the possible role of undetected ferromagnetic impurities such as Fe, Co, Ni, etc. (Janisch2005). Even pure single crystalline sapphire substrates show a ferromagnetic behavior due to iron concentration ranged from 1 to 260 nanogram per centimeter squared (Salcer2007). Impurities can be introduced during the procedure used to fix the substrates to the oven (Golmar2008). Nanoparticles of various sizes and shapes were observed as a result of hydrothermal treatment of cyanometalate polymers (Lefebvre 2008). Silicon has been declared as a magnetic element, (Kopnov 2007). RTFM observed in Co doped ZnO grown on Si (100) has been characterized as coming from Si/SiOx interface. (Yin 2008). It was shown that Si FM is not intrinsic: iron from the pyrex glassware appears on silicon substrate after etching in hot KOH in the form of well-separated ferromagnetic particles. Smoking is a possible source of data contamination to be avoided in magnetism laboratories (Cador 2008). The paper (Garcia 2009) analyses the errors coming from such units of SQUID equipment as polyimide Kapton® tape, gelatin capsules, cotton, plastically deformed straws, anisotropy artifacts coming from irregular distributed impurities.

An important source of mistakes in the interpretation of the experimental data is underestimation of the role of iron in the measured magnetic signal. Usually the authors use the following logics: "Let us start with a *rather unrealistic assumption* that all iron impurities form metallic clusters and all clusters are large enough to behave ferro- or ferri-magnetically…" Indeed, in order to contribute to the total magnetization of the samples, the impurities must interact magnetically. Simple addition of transition metals does not lead to ferromagnetic properties, as no interaction pathway is provided. However, the assumption that all atoms of transition metals form clusters large enough, is *not an unrealistic*. If the sample was prepared at high temperatures, metallic atoms could aggregate during the cooling process with the formation of clusters (Lefebvre 2008 ). For iron superparamagnetic limit is of the order of 150 Å, i.e. about $10^5$ atoms per Fe cluster (Kittel 1946). Normally for smaller particles FM is not expected. However, one must take into account the *effects of the local environment* on the electronic structure and magnetic moments which can be different for various structural forms (Liu 1989). If the iron concentration is small, the absence of superparamagnetic behaviour (Babonneau 2000, Enz 2006, Schwickardi 2006) suggests that impurities either do not contribute to magnetic properties, or their role is far from trivial.

The sample may provide special conditions for the metallic atoms to aggregate. One example is iron in nanographite matrix: there is an accidental matching of the Fe–Fe distance 2.866 Å with that of the C1–C4 distance ~2.842 Å of the hexagonal rings in graphite (Kosugi 2004). However, proximity of carbon generally leads the reduced magnetization of iron (Saito 1997, Host 1998, Fauth 2004). Carbon quenches ferromagnetism of iron in the case of stainless steel.

In bulk Fe, magnetism and structure are strongly dependent. The magnetic moments of free Fe monolayers are theoretically found to be larger than those of the surface with values equal to 3.2 $\mu_B$ for Fe(001) (Freeman 1991). Magnetic moments $\mu(N)$ of iron clusters $\mu$ ($25 \leq N \leq 130$) is $3\mu_B$ per atom, decreasing to the bulk value ($2.2\mu_B$ per atom) near $N = 500$. For all sizes, $\mu$ decreases with increasing temperature, and is approximately constant above a temperature $T_C(N)$. For example, $T_C(130)$ is about 700 K, and $T_C(550)$ is about 550 K ($T_C$ bulk =1043 K). (Billas 1993). This means that *neither the absence of large clusters nor an unusual Curie temperature* can be taken as an evidence of iron-independent magnetism. The enhancement of magnetism at the surface can be qualitatively understood from the analysis ob spin imbalance in the bulk and on the surface.(Fritsche 1987).

Small iron clusters are driven into a nonmagnetic state by the interaction to graphitic surfaces (Fauth 2004). Proximity of carbon generally leads the reduced magnetization of the transition metal clusters (Saito 1997, Host 1998). The reduced magnetism in case of very small clusters is explained by the fact that the transition metal 4s-related density of states is strongly shifted upwards in energy due to the repulsive interaction with the carbon $\pi$ orbitals (Duffy 1998). Nontrivial origin of magnetic behaviour in contaminated carbon-based materials may result from catalytic or template properties of transition metal atoms.

Fe impurities weaken the ferromagnetic behavior in Au by delocating the charge from the surface of the NPs (Crespo 2006). Such magnetic element as cobalt suppresses ferromagnetism in intrinsically magnetic ZnO films (Ghoshai 2008).

## 9. Nontrivial role of transition metals. Charge transfer ferromagnetism

One of the mechanisms of nontrivial role of magnetic metals in d-zero ferromagnetism is contact-induced magnetism which arises when nonmagnetic materials brought in the proximity with magnetic ones (Cerpedes 2004) or thin films in C/Fe multilayer stacks (Mertins 2004).

There are theoretical models which explain the role or iron as just the role of a defect, and magnetic nature of the dopant does not play a role. In case of ferromagnetism of undoped $CaB_6$ iron serves as one of the defects which lead to a partially occupied impurity band which is responsible for RTFM (Edwards and Katsnelson 2006).

A non-trivial role of transition metal impurities in oxide ferromagnetism is proposed in Ref. (Osorio-Gullien 2008) The impurities introduce excess electrons in oxides and either (i) introduce resonant states inside the host conduction band and produce free electrons or (ii) introduce a deep gap state that carries a magnetic moment. The second scenario leads to a ferromagnetic behaviour.

The following arguments have been put forward by (Coey 2008) that magnetism in the dilute magnetic oxide films is not related to the transition metal cations: The oxides are not well crystalline materials, on the contrary, magnetism is governed by the defect structure; high Curie temperatures are incompatible with low concentration of magnetic dopants, and the dopants itself are paramagnetic whereas the whole sample is ferromagnetic. Also, the magnetic semiconductor effects such as Hall, Faraday and Kerr effects are *not* observed. Thus, magnetism in these systems is not due to the ferromagnetically ordered moments of the doping cations mediated by the carriers. The role of transition metal ions is to provide a charge reservoir, and this ability is due to the fact that the cations can exist in two different charge states. "It is therefore the ability of the 3d cations to exhibit *mixed valence*, rather than their possession of a localized moment, which is the key to the magnetism" (Coey 2008). The proposed model of a formation of a defect-based narrow band and tuning the position of the Fermi level by transferred charges was named charge-transfer ferromagnetism.

## 10. Interface magnetism

There is growing experimental evidence that a new type of magnetism has been identified, namely, a magnetism related to surfaces and interfaces of nonmagnetic materials, the "interface magnetism" (Brinkman 2007, Eckstein 2007). Similar effects have been descrifbed for different objects: organic molecules adsorbed on metals (Carmeli 2003) $HfO_2$-coated silicon or sapphire(Ventkatesan 2004) silicon/silicon oxide interfaces (Kopnov 2007) or PbS self-assembled nanoparticles on GaAs (Zakrassov 2008). Au, Ru, Rh, Pd which do not show bulk magnetization, becomes magnetic at the nanoscale. Semiconductor nanoparticles show increasing magnetization for decreasing diameter (Neeleshwar 2005), strong size dependence was found for Ge quantum dots (Liou 2008). Ferromagnetism of a different nature is observed in thin films and nanoparticles capped with organic molecules (Crepso 2004, Yamamoto 2004). From the analysis of the anisotropy of thiol capped gold films, a conclusion is made that the *orbital momentum induced at the surface conduction electrons* is crucial to understand the observed giant anisotropy (Hernnando 2006, Hernando 2006b).

## 11. Conclusions

Due to the present status of researches in this field, the author does not take a liberty to give preference to any of the theories. What is unambiguously clear now is that nanoscale magnetism of otherwise nonmagnetic materials is *sui generic,* i.e. 'outside the family'.

**References**
1. Abraham, D. W., Frank, M. M., and Guha, S. 2005. Absence of magnetism in hafnium oxide films. *Appl. Phys. Lett.* 87. 252502.
2. Affoune, A.M., Prasad, B.L.V., Sato H., Enoki, T. Kaburagi, Y. Hishiyama, Y. 2001. Experimental evidence of a single nano-graphene, Chem. Phys. Lett. 348: 17-20.
3. Akutsu, S, Utsushikawa, Y.. 1999. Magnetic properties and electron spin resonance of carbon fine particles prepared in He plasma, *Mat. Sc. Res. Int*, 5: 110-15.
4. Andersson, O. E., Prasad B. L. V., Sato, H. 1998. Structure and electronic properties of graphite nanoparticles. *Phys. Rev B* 58: 16387-93
5. Ando, K., Saito, H., Jin, Z. W., et al. 2001. Magneto-optical properties of ZnO-based diluted magnetic semiconductors. *J. Appl. Phys.* 89, 7284-6.
6. Andriotis, A. N., Menon, M., R. Sheetz, M. 2005. Are s–p- and d-ferromagnetisms of the same origin? *J. Phys.: Condens.* Matter 17: L35–38 .
7. Antonov, V. E.; Bashkin, I.O.; Khasanov, S. S et al. 2002. Magnetic ordering in hydrofullerite $C_{60}H_{24}$. *J. Alloys and Comp.* 330: 365 – 68.
8. Araujo, M.A.N., Peres, N.M.R. 2006.Weak ferromagnetism and spiral spin structures in honeycomb Hubbard planes, *J. Phys.: Condens. Matter* 18 1769-79.
9. Arcon, D., Jaglicic, Z., Zorko, A. et al. 2006. Origin of magnetic moments in carbon nanofoam, *Phys. Rev. B* 74: 014438.
10. Ata, M.; Machida, M.; Watanabe, H.; Seto, J. 1994. Polymer – $C_{60}$ composite with ferromagnetism. *Jpn. J. Appl. Phys.* 33: 1865 - 1871
11. Azevedo, S., de Paivaa, R., Kaschny, J.R. 2008. Spin polarization in carbon nanostructures with disclinations. *Physics Letters A* 372: 2315 – 18.
12. Babonneau, D., Briatico, J., Petrof, F., et al. 2000. Structural and magnetic properties of $Fe_xC_{1-x}$ nanocomposite thin films. *J. Appl. Phys*. 87: 3432-43.
13. Balasubramanian, G., Neumann, P., Twitchen, D. et al. 2009. Ultralong spin coherence time in isotopically engineered diamond. Nature Mat. advance online publications, NMAT2420.
14. Balents, L., and Varma, C. M. 2000. Ferromagnetism in Doped Excitonic Insulators. *Phys. Rev. Lett.* 84: 1264-7.
15. Banerjee, S., Mandal, M., Gayathri, N., and Sardar, M. 2007. Enhancement of ferromagnetism upon thermal annealing in pure ZnO. *Appl. Phys. Lett.* 91.
16. Barborini, E., Piseri, P., Milani, P, et al. 2002. Negatively curved spongy carbon, *Appl. Phys. Lett*. 81: 3359-61.
17. Barzola-Quiquia, J., Esquinazi, P., Rothermel, M, Spemann, D., Butz, T., Garcia, N. 2007. Experimental evidence for two-dimensional magnetic order in proton bombarded graphite. *Phys. Rev. B* 76: 161403.
18. Barzola-Quiquia, J., Hohne, R., Rothermel, M, Setzer, A, Esquinazi, P., Heera, V, 2008. A comparison of the magnetic properties of proton- and iron-implanted graphite *Eur. Phys. J. B* 61: 127-30.

19. Barzykin, V., and Gorkov, L. P. Ferromagnetism and superstructure in $Ca_{1-x}La_xB_6$. 2000. *Phys. Rev. Lett.* 84: 2207-10.
20. Beltran, J. I., Munoz, M. C., and Hafner, J. 2008. Structural, electronic and magnetic properties of the surfaces of tetragonal and cubic $HfO_2$. New J. Phys. 10: 063031.
21. Bennett, M. C., van Lierop, J., Berkeley, E. M., et al. 2004. Weak ferromagnetism in $CaB_6$. *Phys. Rev. B* 69.
22. Berardan, D., Guilmeau, E., and Pelloquin, D. 2008. Intrinsic magnetic properties of $In_2O_3$ and transition metal-doped-$In_2O_3$. *J. Magn. Magn. Mater.* 320: 983-9.
23. Bhosle, V., and Narayan, J. 2008. Observation of room temperature ferromagnetism in Ga:ZnO: A transition metal free transparent ferromagnetic conductor. *Appl. Phys. Lett.* 93.
24. Bhowmick, S., and Shenoy, V. B. 2008. Edge state magnetism of single layer graphene nanostructures, *J. Chem. Phys.* 128: 244717.
25. Billas, I. M. L., Becker, J. A., Châtelain, A., and de Heer, W. A. 1993. Magnetic moments of iron clusters with 25 to 700 atoms and their dependence on temperature. *Phys. Rev. Lett.* 71: 4067-70.
26. Blinc, R., Cevc, P., Arcon, D., et al. 2006. C-13 NMR and EPR of carbon nanofoam. *phys. stat. sol b* 243: 3069-72
27. Bogani, L., and Wernsdorfer, W. 2008. Molecular spintronics using single-molecule magnets. *Nature Mater.* 7: 179-86.
28. Botello-Mendez, A.R., Lopez-Urias, F., Terrones, M., et al. 2008. Magnetic behavior in zinc oxide zigzag nanoribbons. Nano Lett. 8: 1562-5.
29. Boukhvalov, D. W., Karimov, R. F., Kurmaev, E. Z., et al. 2004. Testing the magnetism of polymerized fullerene. *Phys. Rev. B* 69: 115425.
30. Boukhvalov, D. W., Katsnelson, M. I., Lichtenstein, A. I., 2008. Hydrogen on graphene: Electronic structure, total energy, structural distortions and magnetism from first-principles calculations. *Phys. Rev. B* 77, 035427.
31. Boukhvalov, D. W., Katsnelson, 2008. Chemical functionalizatyion of graphene with defects. *Nano Lett.,* 8, 473-9.
32. Borges, R. P., da Silva, R. C., Magalhaes, S., Cruz, M. M., and Godinho, M. 2007. Magnetism in Ar-implanted ZnO. *J. Phys.: Condens. Matter* 19.
33. Bouzerar, G., and Ziman, T. 2006. Model for vacancy-induced d(0) ferromagnetism in oxide compounds. *Phys. Rev. Lett.* 96.
34. Brinkman, A., Huijben, M., Van Zalk, M., et al. 2007. Magnetic effects at the interface between non-magnetic oxides. *Nature Mater.* 6.
35. Buchholz, D. B., Chang, R. P. H., Song, J. H., and Ketterson, J. B. 2005. : Room-temperature ferromagnetism in Cu-doped ZnO thin films *Appl. Phys. Lett.* 87.
36. Cador, O., Caneschi, A., Rovai, D., Sangregorio, C., Sessoli, R., and Sorace, L. 2008. From multidomain particles to organic radicals: The multifaceted magnetic properties of tobacco and cigarette ash. Inorg. Chim. Acta361: 3882-6.
37. Cahen, D., Naaman, R., and Vager, Z. 2005. The cooperative molecular field effect. Adv. Funct. Mater.15: 1571-8.
38. Calderón, M. J., and Das Sarma, S. 2007. Theory of carrier mediated ferromagnetism in dilute magnetic oxides. Annals of Physics 322: 2618-34.
39. Carmeli, I., Leitus, G., Naaman, R., Reich, S., and Vager, Z. 2003. Magnetism induced by the organization of self-assembled monolayers. *J. Chem. Phys.* 118: 10372-5.
40. Carmeli, I., Bloom, F., Gwinn, E. G., et.al. 2006. Molecular enhancement of ferromagnetism in GaAs/GaMnAs heterostructures. *Appl. Phys. Lett.* 89.
41. Caudillo, R., Gao, X., Escudero, R., et al. 2006. Ferromagnetic behavior of carbon nanospheres encapsulating silver nanoparticles, *Phys. Rev. B* 74: 214418.
42. Céspedes, O., Ferreira, M. S., Sanvito, S., Kociak, M., and Coey, J. M. D. 2004. Contact induced magnetism in carbon nanotubes. *J. Phys.: Condens. Matter* 16: L155–61.
43. Chakraborti, D., Narayan, J., and Prater, J. T. 2007. Room temperature ferromagnetism in $Zn_{1-x}Cu_xO$ thin films *Appl. Phys. Lett.* 90: 062504.
44. Chan, J. A., Montanari, B. Gale, J. D., Bennington, S. M., Taylor, J. W., and Harrison. N. M., 2004. Magnetic properties of polymerized $C_{60}$: The influence of defects and hydrogen. *Phys. Rev B 70,* 041403.
45. Cho, B. K., Rhyee, J. S., Oh, B. H., et al. 2004. Formation of midgap states and ferromagnetism in semiconducting $CaB_6$. *Phys. Rev. B* 69.
46. Cho, J.-H. and Choi, J. – H. 2008. Antiferromagnetic ordering in one-dimensional dangling-bond wires on a hydrogen-terminated C(001) surface: A density-functional study. *Phys. Rev. B* 77: 075404.
47. Choi, J., Kim, Y., Chang, K. J, and Tomanek, D. 2003. Itinerant ferromagnetism in heterostructured C/BN nanotubes. *Phys. Rev. B* 67: 125421.
48. Ceperley, D. 1999. Return of the itinerant electron. *Nature* 397: 386-7.
49. Coey, M., and Sanvito, S. 2004. The magnetism of carbon. *Phys. World* 17: 33-7.
50. Coey, J. M. D., Venkatesan, M., Stamenov, P., Fitzgerald, C. B., and Dorneles, L. S. 2005. Magnetism in hafnium dioxide. *Phys. Rev. B* 72.
51. Coey, J. M. D. d(0) ferromagnetism. 2005. *Solid State Sci.* 7: 660-7.
52. Coey, J. M. D., Venkatesan, M., and Fitzgerald, C. B. 2005c. Donor impurity band exchange in dilute ferromagnetic oxides. *Nature Mater.* 4: 173-9.
53. Coey, J. M. D., Wongsaprom, K., Alaria, J., and Venkatesan, M. 2008. Charge-transfer ferromagnetism in oxide nanoparticles. *J. Phys. D Appl. Phys* 41: 134012.
54. Crespo, P., Garcia, M. A., Pinel, E. F., et al. 2006. Fe impurities weaken the ferromagnetic behavior in Au nanoparticles. *Phys. Rev. Lett.* 97: 177203.
55. Crespo, P., Litran, R., Rojas, T. C., et al. 2004. Permanent magnetism, magnetic anisotropy, and hysteresis of thiol-capped gold nanoparticles. *Phys. Rev. Lett.* 93: 087204.
56. Crespo, P., Garcia, M. A., Fernandez-Pinel, E., et al. 2008. Permanent magnetism in thiol capped nanoparticles, gold and ZnO. *Acta Phys. Pol. A* 113: 515-20.
57. De La Venta, J., Pinel, E. F., Garcia, M. A., Crespo, P., and Hernando, A. 2007. Magnetic properties of organic coated gold surfaces. *Mod. Phys. Lett. B* 21: 303-19.
58. Denlinger, J. D., Clack, J. A., Allen, J. W., et al. 2002. Bulk band gaps in divalent hexaborides. *Phys. Rev. Lett.* 89: 157601.
59. Dev, P., Xue, Y., and Zhang, P. 2008. Defect-induced intrinsic magnetism in wide-gap III nitrides. *Phys. Rev. Lett.*100: 117204.
60. Dietl, T., Ohno, H., Matsukura, F., Cibert, J., and Ferrand, D. 2000. Zener model description of ferromagnetism in zinc-blende magnetic semiconductors. *Science* 287: 1019-22.
61. Dietl, T. 2008. Origin and control of ferromagnetism in dilute magnetic semiconductors and oxides (invited). *J. Appl. Phys.* 103.
62. Dorneles, L. S., Venkatesan, M., Moliner, M., Lunney, J. G., and Coey, J. M. D.. 2004. Magnetism in thin films of $CaB_6$ and $SrB_6$. *Appl. Phys. Lett.* 85: 6377-9.
63. Duffy, D. M., and Blackman, J. A. 1998. Magnetism of 3d transition-metal adatoms and dimers on graphite. *Phys. Rev. B* 58: 7443-9.

64. Durst, A. C., Bhatt, R. N., and Wolff, P. A. 2002. Bound magnetic polaron interactions in insulating doped diluted magnetic semiconductors. *Phys. Rev. B* 65.
65. Dutta, P., Pal, S., Seehra, M. S., et al. 2007. Magnetism in dodecanethiol-capped gold nanoparticles: Role of size and capping agent. *Appl. Phys. Lett.* 90: 213102.
66. Dutta, P., Seehra, M. S., Zhang, Y., and Wender, I. 2008. Nature of magnetism in copper-doped oxides: $ZrO_2$, $TiO_2$, MgO, $SiO_2$, $Al_2O_3$, and ZnO. *J. Appl. Phys.* 103: 07D104.
67. Eckstein, J. N. 2007. Oxide interfaces - Watch out for the lack of oxygen. *Nature Mater.* 6: 473-4.
68. Edwards, D. M., and Katsnelson, M. I. 2006. High-temperature ferromagnetism of sp electrons in narrow impurity bands: application to CaB6. *J. Phys.: Condens. Matter* 18: 7209-25.
69. Elfimov, I. S., Yunoki, S., and Sawatzky, G. A. 2002. Possible path to a new class of ferromagnetic and half-metallic ferromagnetic materials. *Phys. Rev. Lett.* 89.
70. Enoki, T., and Takai, K. 2006. Unconventional magnetic properties of nanographite, in: : *Carbon Based Magnetism*, Ed. by T.L.Makarova and F. Palacio. Elsevier, 397-416.
71. Enoki, T., Kobayashi, Y., Fukui, K.I. 2007. Electronic structures of graphene edges and nanographene. Int. Rev. Phys. Chem. 26: 609-45.
72. Enoki, T., Takai, K. 2008. Unconventional electronic and magnetic functions of nanographene-based host-guest systems. Dalton Trans. 29: 3773 - 81
73. Esquinazi, P., Setzer, A, R. Höhne, R, et al. 2002. Ferromagnetism in oriented graphite samples. *Phys. Rev. B* 66: 24429.
74. Esquinazi, P., Spemann, D., Höhne, R., Setzer, A, Han, K.-H. and Butz, T., 2003. Induced magnetic ordering by proton irradiation in graphite. *Phys. Rev. Lett.* 91: 227201.
75. Faccio, R., Pardo, H., Denis, P.A., et al. 2008. Magnetism induced by single carbon vacancies in a three-dimensional graphitic network. *Phys. Rev. B* 77: 035416.
76. Fauth, K., Gold, S., Hessler, M., Schneider, N., and Schutz, G. 2004. Cluster surface interactions: small Fe clusters driven nonmagnetic on graphite. *Chem. Phys. Lett.* 392: 498-502.
77. Feng Liu., Press, M. R., Khanna, S. N., and Jena, P. 1989. Magnetism and local order: *Ab initio* tight-binding theory. *Phys. Rev. B* 39: 6914 – 24.
78. Fernandez-Rossier, J., Palacios, J.J. 2007. Magnetism in graphene nanoislands *Phys. Rev. Lett.* 99: 177204.
79. Fernandez-Rossier, J., 2008. Prediction of hidden multiferroic order in graphene zigzag ribbons. *Phys. Rev. B*, 77: 075430.
80. Fisk, Z., Ott, H. R., Barzykin, V., et al. 2002. The emerging picture of ferromagnetism in the divalent hexaborides. Physica B 312: 808-10.
81. Freeman, A. J., and Wu, R. Q. 1991. Electronic structure theory of surface, interface and thin-film magnetism. *J. Magn. Magn. Mater.* 100: 497-514.
82. Fritsche, L., Noffke, J., and Eckard, H. 1987. Relativistic treatment of interacting spin-aligned electron systems: application to ferromagnetic iron, nickel and palladium metal. *J. Phys. F* 17: 943-65.
83. Fujita, M., Wakabayashi, K., Nakada, K., and Kusakabe, K., 1996. Peculiar localized state at zigzag graphite edge. *J. Phys. Soc. Jpn.* 65: 1920-23.
84. Gavilano, J. L., Mushkolaj, S., Rau, D., et al. 2001. Anomalous NMR spin-lattice relaxation in SrB6 and Ca1-xLaxB6. *Phys. Rev. B* 63: 140410.
85. Gallego, S., Beltran, J. I., Cerda, J., and Munoz, M. C. 2005. Magnetism and half-metallicity at the O surfaces of ceramic oxides. *J. Phys.: Condens. Matter* 17: L451-7.
86. Garcia, M.A., Pinel, E.F., de la Venta, J. et al. 2009. Sources of experimental errors in the observation of nanoscale magnetism. *J. Appl. Phys.* 105: 013925.
87. Garitaonandia, J. S., Insausti, M., Goikolea, E., et al. 2008. Chemically induced permanent magnetism in Au, Ag, and Cu nanoparticles: Localization of the magnetism by element selective techniques. *Nano Lett.* 8: 661-7.
88. Garcia, M. A., Merino, J. M., Pinel, E. F., et al. 2007. Magnetic properties of ZnO nanoparticles. *Nano Lett.* 7: 1489-94.
89. Ge, M. Y., Wang, H., Liu, E. Z., et al. 2008. On the origin of ferromagnetism in CeO2 nanocubes. *Appl. Phys. Lett.* 93: 062505
90. Gianno, K., Sologubenko, A. V., Ott, H. R., Bianchi, A. D., and Fisk, Z. 2002. Low-temperature thermoelectric power of CaB6. *J. Phys.: Condens. Matter* 14: 1035-43.
91. Golmar, F., Navarro, A. M. M., Torres, C. E. R., et al. 2008. Extrinsic origin of ferromagnetism in single crystalline LaAlO3 substrates and oxide films. *Appl. Phys. Lett.* 92: 262503.
92. Ghoshal, S., and Kumar, P. S. A. 2008. Suppression of the magnetic moment upon Co doping in ZnO thin film with an intrinsic magnetic moment. *J. Phys.: Condens. Matter* 20: 192201.
93. González, J., Guinea, F., and Vozmediano, M.A.H. 2001. Electron-electron interactions in graphene sheets, *Phys. Rev. B* 63: 134421.
94. Grace, P.J., Venkatesan, M., Alaria, J., Coey, J.M.D., Kopnov, G., Naaman, R. 2009. The origin of the magnetism of etched silicon. *Adv. Mat.* 21: 71-4.
95. Guerrero, E., Munoz-Marquez, M. A., Garcia, M. A., et al. 2008. Surface plasmon resonance and magnetism of thiol-capped gold nanoparticles. *Nanotechnology* 19: 17501
96. F. Guinea, F, M. Lopez-Sancho,P, and Vozmediano, M. A. H. 2006. Interactions and disorder in 2D graphite sheets, in: *Carbon based magnetism,* Ed. by T.L.Makarova and F. Palacio. Elsevier, 353-71.
97. Hadacek, N., Nosov, A., Ranno, L., Strobel, P., and Galera, R.-M. 2007. Magnetic properties of HfO2 thin films. *J. Phys.: Condens. Matter* 19: 486206
98. Hassini, A., Sakai, J., Lopez, J. S., and Hong, N. H. 2008. Magnetism in spin-coated pristine TiO2 thin films. *Phys. Lett. A* 372: 3299-302.
99. Han, K.H., Spemann, D., Hohne, R., et al. 2003. Observation of intrinsic magnetic domains in $C_{60}$ polymers. *Carbon* 41, 785 - 95.
100. Heisedberg, W. 1928. Zur Theorie des Ferromagnetismus. *Z. Phys.* 49: 615-36.
101. Hernando, A., Crespo, P., and García, M. A. 2006. Origin of Orbital Ferromagnetism and Giant Magnetic Anisotropy at the Nanoscale. *Phys. Rev. Lett.* 96: 057206.
102. Hernando, A., Crespo, P., García, M. A., et al. 2006. Giant magnetic anisotropy at the nanoscale: Overcoming the superparamagnetic limit. *Phys. Rev. B* 74: 052403.
103. Han, K.H., Spemann, D., Hohne, R., et al. 2003. Observation of intrinsic magnetic domains in $C_{60}$ polymers. *Carbon* 41, 785 - 95.
104. Hong, N. H., Sakai, J., Ruyter, A., and Brize, V. 2006. Does Mn doping play any key role in tailoring the ferromagnetic ordering of TiO2 thin films? *Appl. Phys. Lett.* 89: 252504.
105. Hong, N. H., Sakai, J., Poirot, N., and Brize, V. 2006. Room-temperature ferromagnetism observed in undoped semiconducting and insulating oxide thin films. *Phys. Rev. B* 73 132404.

106. Hong, N. H., Sakai, J., and Gervais, F. 2007. Magnetism due to oxygen vacancies and/or defects in undoped semiconducting and insulating oxide thin films. *J. Magn. Magn. Mater.* 316: 214-7.
107. Hong, N. H., Sakai, J., and Brize, V. 2007. Observation of ferromagnetism at room temperature in ZnO thin films. *J. Phys.: Condens. Matter* 19: 036219.
108. Hong, N. H., Poirot, N., and Sakai, J. 2008. Ferromagnetism observed in pristine SnO2 thin films. *Phys. Rev. B* 77: 033205.
109. Hong, J. S. 2008. Local magnetic moment induced by Ga vacancy defect in GaN. *J. Appl. Phys.* 103: 063907.
110. Hou, D. L., Ye, X. J., Meng, H. J., et al. 2007. Magnetic properties of n-type Cu-doped ZnO thin films. *Appl. Phys. Lett.* 90 142502.
111. Hori, H., Teranishi, T., Nakae, Y., et al. 1999. Anomalous magnetic polarization effect of Pd and Au nano-particles. *Phys. Lett. A* 263: 406-10.
112. Hori, H., Yamamoto, Y., Iwamoto, T., Miura, T., Teranishi, T., and Miyake, M. 2004. Diameter dependence of ferromagnetic spin moment in Au nanocrystals. *Phys. Rev. B* 69: 174411.
113. Host, J. J., Block, J. A., Parvin, K., et al. 1998. Effect of annealing on the structure and magnetic properties of graphite encapsulated nickel and cobalt nanocrystals. *J. Appl. Phys.* 83: 793-801.
114. Huang, L. M., Rosa, A. L., and Ahuja, R. 2006. Ferromagnetism in Cu-doped ZnO from first-principles theory. *Phys. Rev. B* 74: 075206 (2006).
115. Huang, B., Feng Liu, Wu, J., Gu, B.-L. and Duan, W. 2008. Suppression of spin polarization in graphene nanoribbons by edge defects and impurities, *Phys. Rev. B* 77: 153411.
116. Ishii, C., Matsumura, Y., and Kaneko, K., 1995, ferromagnetic behaviour of superhigh surface-area carbon. *J. Phys. Chem.* 99: 5743-5.
117. Ito, Y., Miyazaki A., Fukui F., Valiyaveettil S., Yokoyama T., and Enoki, T. 2008. Pd Nanoparticle Embedded with Only One Co Atom Behaves as a Single-Particle Magnet. *J. Phys. Soc. Jap* 77: 103701.
118. Ivanovskii, A. L. 2007. Magnetic effects induced by sp impurities and defects in nonmagnetic sp materials. *Phys.-Usp.* **50** 1031-52 .
119. Janisch, R., Gopal, P., Spaldin, N.A.2005. Transition metal-doped TiO2 and ZnO - present status of the field. *J. Phys.: Condens. Matter* 17, R657 -89.
120. Jian, W. B., Lu, W. G., Fang, J., Lan, M. D., and Lin, J. J. 2006. Spontaneous magnetization and ferromagnetism in PbSe quantum dots. *J. Appl. Phys.* 99: 08N708.
121. Jungwirth, T., Sinova, J., Masek, J., Kucera, J., and MacDonald, A. H. 2006. Theory of ferromagnetic (III,Mn)V semiconductors. *Rev. Mod. Phys.* 78: 809-64.
122. Katayama-Yoshida, H., Sato, K., Fukushima, T., et al. 2007. Theory of ferromagnetic semiconductors. *Phys. Status Solidi A* 204: 15-32.
123. Katsnelson, M. I., Irkhin, V. Y., Chioncel, L., Lichtenstein, A. I., and de Groot, R. A. 2008. Half-metallic ferromagnets: From band structure to many-body effects. *Rev. Mod. Phys.* 80: 315-78.
124. Kenmochi, K., Seike, M., Sato, K., Yanase, A., and Katayama-Yoshida, H. 2004. New class of diluted ferromagnetic semiconductors based on CaO without transition metal elements, *Jpn. J. Appl. Phys.* 43: L934-6.
125. Keavney, D. J., Buchholz, D. B., Ma, Q., and Chang, R. P. H. 2007. Where does the spin reside in ferromagnetic Cu-doped ZnO? *Appl. Phys. Lett.* 91: 012501.
126. Kim, Y.-H., Choi, J., and Chang. K. J., 2003. Defective fullerenes and nanotubes as molecular magnets: An ab initio study. *Phys. Rev. B* 68: 125420.
127. Kittel, C. 1946. Theory of the Structure of Ferromagnetic Domains in Films and Small Particles. *Phys. Rev.* 70: 965-71.
128. Kittilstved, K. R., Liu, W. K., and Gamelin, D. R. 2006. Electronic structure origins of polarity-dependent high-$T_C$ ferromagnetism in oxide-diluted magnetic semiconductors. *Nature Mater.* 5: 291-7.
129. Kreutz, T. C., Gwinn, E. G., Artzi, R., Naaman, R., Pizem, H., and Sukenik, C. N. 2003. Modification of ferromagnetism in semiconductors by molecular monolayers. *Appl. Phys. Lett.* 83: 4211-3.
130. Kopelevich, Y., Esquinazi, P., Torres, J. H. S., and Moehlecke, S., 2000. *J. Low Temp. Phys.* 119: 691-7.
131. Kopelevich, Y., da Silva, R. R., Torres, J. H. S., Penicaud, A., Kyotani, T. 2003. Local ferromagnetism in microporous carbon with the structural regularity of zeolite. *Phys. Rev. B* 68:092408.
132. Kopelevich, Y., Esquinazi, P. 2007. Ferromagnetism and superconductivity in carbon-based systems. *J. Low Temp. Phys. 146:* 629-39.
133. Kopnov, G., Vager, Z., and Naaman, R. 2007. New magnetic properties of silicon/silicon oxide interfaces. *Adv. Mater.* 19: 925-8.
134. Kosugi, K., Bushiri, M. J., and Nishi, N. 2004. Formation of air stable carbon-skinned iron nanocrystals from FeC2. *Appl. Phys. Lett.* 84 1753-5.
135. Krasheninnikov, A. V., Lehtinen, P.O., Foster, A. S., Pyykko, P. Nieminen, R. M. 2009, Embedding Transition-Metal Atoms in Graphene: Structure, Bonding, and Magnetism. *Phys. Rev.Lett.* 102: 126807.
136. Kumar, A., Avasthi, D. K. Pivin, C., Tripathi, A., Singh, F., 2006. Ferromagnetism induced by heavy-ion irradiation in fullerene films *Phys. Rev. B.* 74: 153409.
137. Kumar, A., Avasthi, D. K. Pivin, C. et al. 2007. Magnetic Force Microscopy of Nano-Size Magnetic Domain Ordering in Heavy Ion Irradiated Fullerene Films. *J. Nanosci. Nanotechno.* 7, 2201–5.
138. Kumazaki, H., Hirashima, D. S. 2007. Possible vacancy-induced magnetism on a half-filled honeycomb lattice. *J. Phys. Soc. Jpn.*, 76: 034707.
139. Kumazaki, H., Hirashima, D. S. 2007b. Nonmagnetic-defect-induced magnetism in graphene. J. Phys. Soc. Jpn., 76, 064713.
140. Kumazaki, H., Hirashima, D. S. 2007. Magnetism of a two-dimensional graphite sheet. *J. Magn. Magn. Mat.* 310: 2256.
141. Kumazaki, H., Hirashima, D. S., 2008. Local magnetic moment formation on edges of graphene. *J. Phys. Soc. Japan*.77: 044705.
142. Kunii, S. 2000. Surface-layer ferromagnetism and strong surface anisotropy in Ca/sub 1-x/La/sub x/B/sub 6/ (x=0.005) evidenced by ferromagnetic resonance. *J. Phys. Soc. Jap.* 69: 3789-91.
143. Kusakabe, K. and Maruyama, M., 2003. Magnetic nanographite. *Phys. Rev. B* 67: 092406.
144. Kusakabe, K., Geshi, M., Tsukamoto, H., and Suzuki, N. 2004. Design of new ferromagnetic materials with high spin moments by first-principles calculation. *J. Phys.: Condens. Matter* 16: S5639-44.
145. Kusakabe, K. 2006. Flat-band ferromagnetism in Organic Crystals, in: *Carbon based magnetism*, Ed. by T.L.Makarova and F. Palacio. Elsevier, 305-28.
146. Kvyatkovskii, O. E., I.B. Zakharova, A.L. Shelankov, and Makarova, T. L. 2004. Electronic Properties of the $(C_{60})_2$ and $(C_{60})_2^{2-}$ Fullerene Dimer. *AIP Conf. Proc.* 723: 385-8.
147. Kvyatkovskii, O. E., Zakharova, I. B. , Shelankov, A. L. , Makarova, T. L. 2005. Spin-transfer mechanism of ferromagnetism in polymerized fullerenes: Ab initio calculations *Phys. Rev. B* 72: 214426.
148. Kvyatkovskii, O. E., Zakharova, I. B. , Shelankov, A. L. 2006. Magnetic Properties of Polymerized Fullerene Doped with Hydrogen, Fluorine and Oxygen. *Fuller., Nanot., Car. Nan.* 14: 373-80.
149. Lei, Y., Shevlin, S. A. Zhu, W., and Guo, Z. X. 2008. Hydrogen-induced magnetization and tunable hydrogen storage in graphitic structures *Phys. Rev. B* 77: 134114.

150. Lehtinen, P.O., Foster, A.S., Ayuela, A., Krasheninnikov, A., Nordlund, K., Nieminen. R.M., 2003. Magnetic properties and diffusion of adatoms on a graphene sheet. *Phys. Rev. Lett.* 91: 017202.
151. Lefebvre, J., Trudel, S., Hill, R. H., et al. 2008. A closer look: Magnetic behavior of a three-dimensional cyanometalate coordination polymer dominated by a trace amount of nanoparticle impurity. *Chem – Eur. J*. 14: 7156-67
152. Lee, J. Y., Cho, J.-H., Kang, M. H. 2009, Antiferromagnetic Ground State of a C60-Covered Si(001) Surface, ChemPhysChem 10, 334 – 6
153. Li, S. D., Huang, Z.G., Lue, L.Y., et al. 2007. Ferromagnetic chaoite macrotubes prepared at low temperature and pressure. *Appl. Phys. Lett.* 90: 232507.
154. Li, D., Han, Z., Wu, B., et al. 2008 a. Ferromagnetic and spin-glass behaviour of nanosized oriented pyrolytic graphite in Pb-C nanocomposites. *J. Phys. D Appl. Phys* 41: 115005.
155. Li, Q. K., Wang, B., Woo, C. H., Wang, H., Zhu, Z. Y., and Wang, R. 2008 b. Origin of unexpected magnetism in Cu-doped $TiO_2$. *Europhys. Lett.*, 81: 17004.
156. Liou, Y., Shen, Y.L., 2008. Magnetic properties of germanium quantum dots. Adv. Mat. 20: 779-83.
157. Liu, X., Bauer, M., Bertagnolli, H., Roduner, E., van Slageren, J., and Phillipp, F. 2006. Structure and Magnetization of Small Monodisperse Platinum Clusters. *Phys. Rev. Lett.* 97: 253401.
158. Liu, Y. L., Lockman, Z., Aziz, A., and MacManus-Driscoll, J. 2008. Size dependent ferromagnetism in cerium oxide ($CeO_2$) nanostructures independent of oxygen vacancies. *J. Phys.: Condens. Matter* 20: 165201.
159. Lobach, A. S.; Shul'ga, Y. M.; Roshchupkina, O. S. et al. 1998. $C_{60}H_{18}$, $C_{60}H_{36}$ and $C_{70}H_{36}$ fullerene hydrides: Study by methods of IR, NMR, XPS, EELS and magnetochemistry. *Fullerene Sci. Technol.* 6: 375-91.
160. Lofland, S. E., Seaman, B., Ramanujachary, K. V., Hur, N., and Cheong, S. W. 2003. Defect driven magnetism in calcium hexaboride. *Phys. Rev. B* 67. 020410.
161. Lounis, S., Dederichs, P. H., and Bluegel, S. 2008. Magnetism of nanowires driven by novel even-odd effects. *Phys. Rev. Lett.* 101: 107204.
162. Ma, Y. C., Lehtinen, P. O., Foster, A. S., and Nieminen, R. M. 2004. Magnetic properties of vacancies in graphene and single-walled carbon nanotubes. *New J. Phys.* 6: 68.
163. Ma, Y. C., Lehtinen, P. O., Foster, A. S., and Nieminen, R. M, 2005. Hydrogen-induced magnetism in carbon nanotubes, *Phys. Rev. B* 72: 085451.
164. Ma, Y. W., Yi, J. B., Ding, J., Van, L. H., Zhang, H. T., and Ng, C. M. 2008. Inducing ferromagnetism in ZnO through doping of nonmagnetic elements. *Appl. Phys. Lett.* 93: 042514.
165. Maca, F., Kudrnovsky, J., Drchal, V., and Bouzerar, G. 2008. Magnetism without magnetic impurities in $ZrO_2$ oxide. *Appl. Phys. Lett.* 92: 212503.
166. Mackay, A. L., Terrones, H., 1991. Diamond from graphite. Nature 352, 762-2
167. Madhu, C., Sundaresan, A., and Rao, C. N. R. 2008. Room-temperature ferromagnetism in undoped GaN and CdS semiconductor nanoparticles. *Phys. Rev. B* 77: 201306.
168. Maiti, K. 2008. Role of vacancies and impurities in the ferromagnetism of semiconducting CaB6. *Europhys. Lett.* 82: 67006.
169. Maiti, K., Medicherla, V. R. R., Patil, S., and Singh, R. S. 2007. Revelation of the role of impurities and conduction electron density in the high resolution photoemission study of ferromagnetic hexaborides. *Phys. Rev. Lett.* 99: 266401.
170. Makarova, T. L., Sundqvist, B, Höhne, R. et al. 2001, Magnetic carbon. *Nature* 413: 716-9.
171. Makarova, T. L., Han, K.-H., Esquinazi, P., et al 2003. Magnetism in photopolymerized fullerenes. *Carbon* 41 (8) 1575 - 1584 2003.
172. Makarova, T. L., Sundqvist, B, Höhne, R. et al. 2001, Magnetic carbon. (Retraction). *Nature* 440: 707-7.
173. Carbon Based Magnetism: An Overview of the Magnetism of Metal Free Carbon-Based Compounds and Materials *edited by T. Makarova and F. Palacio* (Elsevier, 2006).
174. Makarova, T. L., Zakharova, I. B. 2008, Separation of intrinsic and extrinsic contribution to fullerene magnetism, *Fuller., Nanot., Car. Nan.* 16: 567-73.
175. Massidda, S., Continenza, A., de Pascale, T. M., and Monnier, R. 1997. Electronic structure of divalent hexaborides. *Z. Phys. B: Condens. Matter* 102: 83-9.
176. Mathew, S., Satpati, B, Joseph B. et al. 2007. Magnetism in $C_{60}$ films induced by proton irradiation *Phys. Rev. B* 75: 075426.
177. Matsubayashi, K., Maki, M., Tsuzuki, T., Nishioka, T., and Sato, N. K. 2002. Magnetic properties - Parasitic ferromagnetism in a hexaboride? *Nature* 420: 143-4.
178. Matsubayashi, K., Maki, M., Moriwaka, T., et al. 2003. Extrinsic origin of high-temperature ferromagnetism in CaB6. *J. Phys. Soc. Jap.* 72: 2097-102.
179. Matsumoto, Y., Murakami, M., Shono, T., et al. 2001. Room-temperature ferromagnetism in transparent transition metal-doped titanium dioxide. *Science* 291: 854-6.
180. Meegoda, C., Trenary, M., Mori, T., and Otani, S. 2003. Depth profile of iron in a $CaB_6$ crystal. *Phys. Rev. B* 67: 172410.
181. Medicherla, V. R. R., Patil, S., Singh, R. S., and Maiti, K. 2007. Origin of ground state anomaly in LaB6 at low temperatures. *Appl. Phys. Lett.* 90: 062507.
182. Meier, R. J., and Helmholdt, R. B. 1984. Neutron-diffraction study of α- and β-oxygen. *Phys. Rev. B* 29: 1387-93.
183. Mencken, H. L. 1920. Prejudices: Second Series.
184. Mertins, H.-C., Valencia, S., Gudat, W., Oppeneer, P. M., Zaharko, O., and Grimmer, H. 2004. Direct observation of local ferromagnetism on carbon in C/Fe multilayers. *Europhys. Lett.* 66: 743-8.
185. Michael, F., Gonzalez, C., Mujica, V., Marquez, M., and Ratner, M. A. 2007. Size dependence of ferromagnetism in gold nanoparticles: Mean field results. *Phys. Rev. B* 76: 224409.
186. Min, H., Borghi, G. Polini, M. and MacDonald, A. H. 2008. Pseudospin magnetism in graphene. *Phys. Rev. B* 77, 041407(R).
187. Mombru, A. W., Pardo, H., Faccio, R., et al. 2005. Multilevel ferromagnetic behavior of room-temperature bulk magnetic graphite. Phys. Rev. B 71:100404.
188. Monnier, R., and Delley, B. 2001. Point Defects, Ferromagnetism, and Transport in Calcium Hexaboride. *Phys. Rev. Lett.* 87: 157204.
189. Mori, T., and Otani, S. 2002. Ferromagnetism in lanthanum doped CaB6: is it intrinsic? *Solid State Commun.* 123: 287-90.
190. Murakami, Y and Suematsu, H. 1996. Magnetism of $C_{60}$ induced by photo-assisted oxidation. *Pure Appl. Chem.* 68: 1463-7.
191. Murakami, S., Shindou, R., Nagaosa, N., and Mishchenko, A. S. 2002. Theory of Ferromagnetism in $Ca_{1-x}La_xB_6$. *Phys. Rev. Lett.* 88: 126404.
192. Naaman, R., and Vager, Z. 2006. New electronic and magnetic properties emerging from adsorption of organized organic layers. *Phys. Chem. Chem. Phys.* 8: 2217-24.
193. Nakano, T., Ikemoto, Y., and Nozue, Y. 2000. Loading density dependence of ferromagnetic properties in potassium clusters arrayed in a simple cubic structure in zeolite LTA. *Physica B* 281: 688-90.

194. Nakano, S., Kitagawa, Y., Kawakami, T., Okumura, M., Nagao, H., Yamaguchi, K. 2004.Theoretical studies on electronic states of Rh-C$_{60}$. Possibility of a room-temperature organic ferromagnet. *Molecules* 9: 792-807.
195. Nakano, T., Gotoa, K., Watanabeb, I., Prattc, F.L., Ikemotod, Y., and Nozue, Y. 2006. μSR study on ferrimagnetic properties of potassium clusters incorporated into low silica X zeolite. *Physica B* 374: 21-5.
196. Narozhnyi, V. N., Müller, K.-H., Eckert, D., et al. 2003. Ferromagnetic carbon with enhanced Curie temperuture. *Physica B*, 329: 1217 - 8.
197. Neeleshwar, S., Chen, C. L., Tsai, C. B., Chen, Y. Y., and Chen, C. C. 2005. Size-dependent properties of CdSe quantum dots. *Phys. Rev. B* 71 : 201307.
198. Negishi, Y., Tsunoyama, H., Suzuki, M., et al. 2006. X-ray magnetic circular dichroism of size-selected, thiolated gold clusters. *J. Am. Chem. Soc.* 128: 12034-5.
199. Norton, D. P., Overberg, M. E., Pearton, S. J., et al. 2003. Ferromagnetism in cobalt-implanted ZnO. *Appl. Phys. Lett.* 83: 5488-90.
200. Nozue, Y., Kodaira, T., and Goto, T. 1992. Ferromagnetism of potassium clusters incorporated into zeolite LTA. *Phys. Rev. Lett.* 68: 3789-92.
201. Ohldag, H., Tyliszczak, T., Hohne, R, et al., 2007. *Phys. Rev. Lett.* 98:187204.
202. Ohno, H. 1998. Making nonmagnetic semiconductors ferromagnetic. *Science* 281: 951-6.
203. Okada, S. and Oshiyama, A., 2001. Magnetic ordering in hexagonally bonded sheets with first-row elements. *Phys. Rev. Lett.* 87, 146803.
204. Okada, S. and Oshiyama, A.,.2003. Electronic structure of metallic rhombohedral C-60 polymers *Phys. Rev. B* 68: 235402.
205. Osorio-Guillen, J., Lany, S., Barabash, S. V., and Zunger, A. 2007. Nonstoichiometry as a source of magnetism in otherwise nonmagnetic oxides: Magnetically interacting cation vacancies and their percolation. *Phys. Rev. B* 75: 184421.
206. Osorio-Guillen, J., Lany, S., and Zunger, A. 2008. Atomic control of conductivity versus ferromagnetism in wide-gap oxides via selective doping: V, Nb, Ta in anatase TiO2. *Phys. Rev. Lett.* 100: 036601.
207. Oshiyama, A., and Okada, S. 2006. Magnetism in Nanometer-scale Materials that Contain No Magnetic Elements. In *Carbon Based Magnetism,* ed. T. Makarova and F. Palacio. Elsevier.
208. Osipov, V., Baidakova, M., Takai, K., Enoki, T., Vul', A. 2006. Magnetic properties of hydrogen-terminated surface layer of diamond nanoparticles. *Fuller. Nanot. Car. Nan.* 14 565-72.
209. Palacios, J. J. and Fernández –Rossier, L. 2008. Vacancy-induced magnetism in graphene and graphene ribbons, *Phys. Rev. B* 77:195428.
210. Otani, S., and Mori, T. 2002. Flux growth and magnetic properties of CaB6 crystals. *J. Phys. Soc. Jap.* 71: 1791-2.
211. Otani, S., and Mori, T. 2003. Flux growth of CaB$_6$ crystals. *J. Mater. Sci.* 22: 1065-6.
212. Ott, H.R., Gavilano, J.L., Ambrosini, B., et al. 2000. Unusual magnetism of hexaborides. *Physica B* 281: 423-7.
213. Owens, F. J., Iqbal, Z., Belova, L., et al. 2004. Evidence for high-temperature ferromagnetism in photolyzed C$_{60}$. *Phys. Rev. B* 69: 033403.
214. Das Pemmaraju, C., and Sanvito, S. 2005. Ferromagnetism driven by intrinsic point defects in HfO$_2$. *Phys. Rev. Lett.* 94.
215. H. Pan, J. B. Yi, L. Shen, R. Q. Wu, J. H. Yang. J. Y. Lin, Y. P. Feng, J. Ding, L. H. Van, J. H. Yin, Room-Temperature Ferromagnetism in Carbon-Doped ZnO Phys. Rev. Lett. 99, 127201 (2007)
216. Pardo, H., Faccio, R., Araujo-Moreira, F. M., et al. 2006. Synthesis and characterization of stable room temperature bulk ferromagnetic graphite. *Carbon* 44: 565-9.
217. Park, M. S., and Min, B. I. 2003. Ferromagnetism in ZnO codoped with transition metals: Zn1-x(FeCo)(x)O and Zn1-x(FeCu)(x)O. *Phys. Rev. B* 68, 224436.
218. Park, N., Yoon, M., Berber, S., Ihm, J., Osawa, E., and Tomanek, D. 2003. Magnetism in all-carbon nanostructures with negative Gaussian curvature. *Phys. Rev. Lett.* 91 237204.
219. Parkansky. N., Alterkop, B., Boman, R.L., et al. 2008. Magnetic properties of carbon nano-particles produced by a pulsed arc submerged in ethanol. *Carbon* 46: 215-9.
220. Pei, X. Y., Yang, X. P., and Dong, J. M. 2006. Effects of different hydrogen distributions on the magnetic properties of hydrogenated single-walled carbon nanotubes. *Phys. Rev. B* 73: 195417.
221. Peng, H., Li. J., Li., S.-S., Xia, J.-B. 2009. Possible origin of ferromagnetism in undoped anatase TiO2. *Phys. Rev. B* 79: 092411
222. Peres, N. M. R., Guinea, F., and Castro, A. H. 2006. Electronic properties of disordered two-dimensional carbon. *Phys. Rev. B* 73: 125411.
223. Pisani, L., Montanari, B., Harrison, N. M. 2008. A defective graphene phase predicted to be a room temperature ferromagnetic semiconductor. *New J. Phys.* 10, 033002.
224. Potzger, K., Zhou, S.Q., Grenzer, J., et al. 2008. An easy mechanical way to create ferromagnetic defective ZnO. *Appl. Phys. Lett.* 92: 182504.
225. Quesada, A., Garcia, M. A, de la Venta, J., Pinel, E. F., Merino, J. M., and Hernando, A. 2007. Ferromagnetic behaviour in semiconductors: a new magnetism in search of spintronic materials. *Eur. Phys. J. B* 59, 457-61.
226. Qiu, X. Y., Liu, Q. M., Gao, F., Lu, L. Y., and Liu, J.-M. 2006. Room-temperature weak ferromagnetism of amorphous HfAlO$_x$ thin films deposited by pulsed laser deposition. *Appl. Phys. Lett.* 89.
227. Rao, M. S. R., Kundaliya, D. C., Ogale, S. B., et al. 2006. Search for ferromagnetism in undoped and cobalt-doped HfO2-delta. *Appl. Phys. Lett.* 88: 142505.
228. Ray, S. G., Daube, S. S., Leitus, G., Vager, Z., and Naaman, R. 2006. Chirality-induced spin-selective properties of self-assembled monolayers of DNA on gold. *Phys. Rev. Lett.* 96. 036101.
229. Reich, S., Leitus, G., and Feldman, Y. 2006. Observation of magnetism in Au thin films. *Appl. Phys. Lett.* 88: 222502.
230. Rhyee, J. S., and Cho, B. K. 2004. The effect of boron purity on electric and magnetic properties of CaB6. *J. Appl. Phys.* 95: 6675-7.
231. Ribas-Arino, J. and Novoa, J. J. 2004. Magnetism in compressed fullerenes. The origin of the magnetic moments in compressed crystals of polymeric C$_{60}$. *Angew. Chem. Int. Ed,* 43: 577 –80.
232. Ribas-Arino, J. and Novoa, J. J. 2004b. Evaluation of the capability of C-60-fullerene to act as a magnetic coupling unit. *J. Phys. Chem. Solids* 65: 787-91.
233. Rode, A. V., Gamaly, E.G., Christy, A.G., et al. 2004. Unconventional magnetism in all-carbon nanofoam. *Phys. Rev. B* 70: 054407.
234. Sahu, B, Min, H., MacDonald, A. H. and Banerjee, S. K. 2008. Energy gaps, magnetism, and electric-field effects in bilayer graphene nanoribbons, *Phys. Rev. B* 78: 045404.
235. Saito, Y., Ma, J., Nakashima, J., and Masuda, M. 1997. Synthesis, crystal structures and magnetic properties of Co particles encapsulated in carbon nanocapsules. *Z. Phys. D* 40: 170-2.
236. Salzer, R., Spemann, D., Esquinazi, P., et al. 2007. Possible pitfalls in search of magnetic order in thin films deposited on single crystalline sapphire substrates. *J. Magn. Magn. Mater.* 317: 53-60.
237. Salamon, M. B., and Jaime, M. 2001. The physics of manganites: Structure and transport. *Rev. Mod. Phys.* 73: 583-628.

238. Sampedro, B., Crespo, P., Hernando, A., et al. 2003. Ferromagnetism in fcc twinned 2.4 nm size Pd nanoparticles. *Phys. Rev. Lett.* 91: 237203.
239. Sanyal, D., Chakrabarti, M., Roy, T.K., and Chakrabarti, A. 2007. The origin of ferromagnetism and defect-magnetization correlation in nanocrystalline ZnO. *Phys. Lett. A* 371: 482-5.
240. Sanchez, N., Gallego, S., and Munoz, M. C. 2008. Magnetic states at the oxygen surfaces of ZnO and Co-doped ZnO. *Phys. Rev. Lett.* 101: 067206.
241. Sato, H., Kawatsu, N., Enoki, T., Endo, M., Kobori, R., Maruyama, S., and Kaneko, K., 2003. Physisorption-induced change in the magnetism of microporous carbon .*Solid State Commun.* 125: 641.
242. Sato, H., Kawatsu, N., Enoki, T., et al. 2007. Physisorption-induced change in the magnetism of microporous carbon. *Carbon* 45:214-7.
243. Shibayama, Y., Sato, H., Enoki, T., et al. 2000. Novel electronic properties of a nano-graphite disordered network and their iodine doping effects. J. Phys. Soc. Jpn. 69: 754-67.
244. Schneider, A. M. Nanomagnetism: A matter of orientation. 2008. Nature Physics 4, 831 - 832
245. Schwickardi, M. , Olejnik, S., Salabas, E. L., Schmidt, W., Schuth, F., 2006. Scalable synthesis of activated carbon with superparamagnetic properties. *Chem. Commun*, 38, 3987-9.
246. Seehra, M. S., Dutta, P., Singh, V., Zhang, Y., and Wender, I. 2007. Evidence for room temperature ferromagnetism in $Cu_xZn_{1-x}O$ from magnetic studies in $Cu_xZn_{1-x}O$/CuO composite. *J. Appl. Phys.* 101: 09H107.
247. Seehra, M. S., Dutta, P., Neeleshwar, S., et al. 2008. Size-controlled Ex-nihilo ferromagnetism in capped CdSe quantum dots. *Adv. Mater.* 20: 1656-60.
248. Shen, L., Wu, R. Q., Pan, H., et al. 2008. Mechanism of ferromagnetism in nitrogen-doped ZnO: First-principle calculations. *Phys. Rev. B* 78: 073306.
249. Shinde, S. R., Ogale, S. B., Higgins, J. S., et al. 2004. Co-occurrence of superparamagnetism and anomalous Hall effect in highly reduced cobalt-doped rutile $TiO_{2-\delta}$ films. *Phys. Rev. Lett.* 92: 166601.
250. Shuai, M., Liao, L., Lu, H.B., et al. 2007. Room-temperature ferromagnetism in Cu+ implanted ZnO nanowires. *J. Phys. D. Appl. Phys.* 41: 135010.
251. Skomski, R. and Coey, J.M.D. 1995. *Permanent Magnetism*. IOP Publishing Bristol.
252. Smogunov, A., Dal Corso, A., Delin, A., Weht, R., and Tosatti, E. 2008. Colossal magnetic anisotropy of monatomic free and deposited platinum nanowires. *Nature Nanotech.* 3: 22-5.
253. Spemann, D. Han, K.H., Hohne R. et al. 2003. Evidence for intrinsic weak ferromagnetism in a $C_{60}$ polymer by PIXE and MFM. *Nucl. Instrum. Meth.* B 210: 531-6.
254. Stauber, T., Guinea, F, and Vozmediano, M.A.H.. 2005. Disorder and interaction effects in two dimensional graphene sheets. *Phys. Rev. B*. 71: 041406.
255. Stauber, T., Castro, E. V., Silva, N. A. P., et al. 2008. First-order ferromagnetic phase transition in the low electronic density regime of biased graphene bilayers, *J. Phys.: Condens. Matter* 20: 335207.
256. Suda, M., Kameyama, N., Suzuki, M., Kawamura, N., and Einaga, Y. 2008. Reversible phototuning of ferromagnetism at Au-S interfaces at room temperature. *Angew. Chem. Intl. Edit.* 47: 160-3.
257. Sudakar C, Kharel P, Suryanarayanan R, et al. 2008. Room temperature ferromagnetism in vacuum-annealed TiO2 thin films. *J. Magn. Magn. Mater.* 320: L31-6.
258. Sundaresan, A., Bhargavi, R., Rangarajan, N., Siddesh, U., and Rao, C.N.R. 2006. Ferromagnetism as a universal feature of nanoparticles of the otherwise nonmagnetic oxides. *Phys. Rev. B* 74: 161306(R).
259. Takai, K., Eto, S., Inaguma, M., and Enoki, T. 2008. Magnetic potassium clusters in the nanographite-based nanoporous system. *J. Phys. Chem. Sol.* 69: 1182-4.
260. Teng, X. W., Han, W. Q., Ku, W., et al. 2008. Synthesis of ultrathin palladium and platinum nanowires and a study of their magnetic properties *Angew. Chem. Intl. Edit* 47: 2055-8.
261. Terashima, T., Terakura, C., Umeda, Y., Kimura, N., Aoki, H., and Kunii, S. 2000. Ferromagnetism vs Paramagnetism and False Quantum Oscillations in Lanthanum-Doped CaB6. *J. Phys. Soc. Jap.* 69: 2423-6.
262. Tietze, T., Gacic, M., Schutz, G., Jakob, G., Bruck, S., and Goering, E. 2008. XMCD studies on Co and Li doped ZnO magnetic semiconductors. *New J. Phys.* 10: 055009.
263. Tirosh, E., and Markovich, G. 2007. Control of defects and magnetic properties in colloidal HfO2 nanorods. *Adv. Mater.* 19: 2608-12.
264. Topsakal, M., Sevinçli, H., and Ciraci1, S., 2008. Spin confinement in the superlattices of graphene ribbons, *Appl. Phys. Lett.* 92, 173118.
265. Tromp, H. J., van Gelderen, P., Kelly, P. J., Brocks, G., and Bobbert, P. A. 2001. CaB6: A new semiconducting material for spin electronics. *Phys. Rev. Lett.* 87: 016401.
266. Ueda, K., Tabata, H., and Kawai, T. 2001. Magnetic and electric properties of transition-metal-doped ZnO films. *Appl. Phys. Lett.* 79: 988-90.
267. Vager, Z., Carmeli, I., Leitus, G., Reich, S., and Naaman, R. 2004. Surprising electronic-magnetic properties of closed packed organized organic layers. *J. Phys. Chem. Sol.* 65: 713-7.
268. Veillette, M. Y., and Balents, L. 2002. Weak ferromagnetism and excitonic condensates. *Phys. Rev. B* 65 : 14428.
269. Venkatesan, M., Fitzgerald, C. B., and Coey, J. M. D. 2004. Unexpected magnetism in a dielectric oxide. *Nature* 430: 630.
270. Vonlanthen, P., Felder, E., Degiorgi, L., et al. 2000. Electronic transport and thermal and optical properties of Ca1-xLaxB6. *Phys. Rev. B*, 62: 10076-82.
271. Vozmediano, M. A. H., Guinea, F., Lopez-Sancho M. P. 2006. Interactions, disorder and local defects in graphite *J. Phys. Chem. Sol.* 67, 562.
272. Wang, X., Liu Z.X., Zhang Y.L., Li F.Y., and Jin C.Q. 2002. Evolution of magnetic behaviour in the graphitization process of glassy carbon. *J. Phys.: Condens. Matter.* 14: 10265.
273. Wang, W. D., Hong, Y. J., Yu, M. H., Rout, B., Glass, G. A., and Tang, J. K. 2006. Structure and magnetic properties of pure and Gd-doped HfO2 thin films. *J. Appl. Phys.* 99: 08M117.
274. Wang, W. C., Kong, Y., He, X., and Liu, B. X. 2006. Observation of magnetism in the nanoscale amorphous ruthenium clusters prepared by ion beam mixing. *Appl. Phys. Lett.* 89: 262511.
275. Wang, Q., Sun, Q., Chen G., Kawazoe, Y., and Jena, P. 2008. Vacancy-induced magnetism in ZnO thin films and nanowires Phys. Rev. B 77, 205411
276. Wang W. L., Meng, S., Kaxiras, E., 2008b. Graphene nanoflakes with large spin. *Nano Lett.* 8: 241-5.
277. Wang, Y., Huang, Y., Song, Y., et al. 2009. Room-Temperature Ferromagnetism of Graphene. Nano Lett. 9: 220-4.
278. Weyer, G., Gunnlaugsson, H. P., Mantovan, R., et al. 2007. Defect-related local magnetism at dilute Fe atoms in ion-implanted. *J. Appl. Phys.* 102: 113915.

279. Wood, R. A., Lewis, M. H., Lees, M. R., et al.2002. Ferromagnetic fullerene, *J. Phys.: Condens. Matter* 14: L385-91.
280. Wu, R. Q., Peng, G. W., Liu, L., Feng, Y. P., Huang, Z. G., and Wu, Q. Y. 2006. Cu-doped GaN: A dilute magnetic semiconductor from first-principles study. *Appl. Phys. Lett.* 89: 062505.
281. Wu. J., Hadelberg., F. 2009. Magnetism in finite-sized single-walled carbon nanotubes of the zigzag type. *Phys. Rev.B 79, 115436.*
282. Xing, G. Z., Yi, J. B., Tao, J. G., et al. 2008. Comparative study of room-temperature ferromagnetism in Cu-doped ZnO nanowires enhanced by structural inhomogeneity. *Adv. Mater.* 20: 3521-7.
283. Xu, Q. Y., Schmidt, H., Zhou, S. Q., et al. 2008. Room temperature ferromagnetism in ZnO films due to defects. *Appl. Phys. Lett.* 92: 082508.
284. Yamamoto Y, Miura T, Suzuki M, et al. 2004. Direct observation of ferromagnetic spin polarization in gold nanoparticles. *Phys. Rev. Lett.* 93: 116801.
285. Yan, Z., Ma, Y., Wang, D., et al. 2008. Impact of annealing on morphology and ferromagnetism of ZnO nanorods. *Appl. Phys. Lett.* 92: 081911.
286. Yazyev, O. V., Helm, L., 2007. Defect-induced magnetism in graphene. *Phys. Rev. B* 75: 125408.
287. Yazyev, O. V., Katsnelson M. I., 2008. Magnetic correlations at graphene edges: Basis for novel spintronics devices. *Phys. Rev. Lett.* 100, 047209.
288. Yazyev, O. V., Wang W. L., Meng, S., Kaxiras, E., 2008. Comment on graphene nanoflakes with large spin: Broken-symmetry states. *Nano Lett.* 9, 766-7.
289. Yazyev, O. V. 2008. Magnetism in disordered graphene and irradiated graphite. *Phys. Rev. Lett.*101: 037203
290. Ye, L.-H., Freeman, A. J., and Delley, B. 2006. Half-metallic ferromagnetism in Cu-doped ZnO: Density functional calculations. *Phys. Rev. B* 73: 033203.
291. Yi, J. B., Pan, H., Lin J. Y., et al. 2008. Ferromagnetism in ZnO nanowires derived from electro-deposition on AAO template and subsequent oxidation. *Adv. Mater.* 20: 1170-4.
292. Yin, Z. G., Chen, N. F., Li, Y., et al. 2008. Interface as the origin of ferromagnetism in cobalt doped ZnO film grown on silicon substrate. *Appl. Phys. Lett.* 93: 142109.
293. Yoon, S. D., Chen, Y., Yang, A., et al. 2006. Oxygen-defect-induced magnetism to 880 K in semiconducting anatase $TiO_2$-delta films. *J. Phys.: Condens.* 18: L355-61.
294. Young, D. P., Hall, D., Torelli, M. E., et al. 1999. High-temperature weak ferromagnetism in a low-density free-electron gas. *Nature* 397: 412-4.
295. Young, D. P., Fisk, Z., Thompson, J. D., Ott, H. R., Oseroff, S. B., and Goodrich, R. G. 2002. Magnetic properties - Parasitic ferromagnetism in a hexaboride? *Nature* 420: 144-4.
296. Yu, D., Lupton, E. M. Liu., M., Feng Liu. 2008. Collective Magnetic Behavior of Graphene Nanohole Superlattices. Nano Res 1: 56 62.
297. Zakrassov A., Leitus G., Cohen S. R., and Naaman, R. 2008. Adsorption-induced magnetization of PbS self-assembled nanoparticles on GaAs. *Adv. Mater.* 20: 2552-5.
298. Zhang, Y., Talapatra, S., , S Kar, Vajtai, R., Nayak, Ajayan, P.M. 2007. First-principles study of defect-induced magnetism in carbon *Phys. Rev. Lett.*, 99:107201.
299. Zhao, Q., Wu, P., Li, B. L., Lu Z. M., and Jiang E. Y. 2008. Room-temperature ferromagnetism in semiconducting $TiO_2$-delta nanoparticles. Chinese Phys. lett. 25: 1811-4.
300. Zhou, S. Q., Potzger, K., Talut, G., et al. 2008. Ferromagnetism and suppression of metallic clusters in Fe implanted ZnO - a phenomenon related to defects? *J. Phys. D.* 41: 105011.
301. Zhou, S, Cizmar, E, Potzger, K. et al. 2009. Origin of magnetic moments in defective $TiO_2$ single crystals. *Phys. Rev. B.*, 79:113201.
302. Zhitomirskyi,M. E., Rice, T. M., and Anisimov, V. I. 1999. Magnetic properties: Ferromagnetism in the hexaborides. *Nature* 402, 251-3.
303. Zorko, A. , Makarova, T. L., Davydov. V. A., et al. 2005. Study of defects in polymerized $C_{60}$: A room-temperature ferromagnet *AIP Conf. Proc.* 786: 21- 4